# On the calibration of stochastic car following models

Shirui Zhou[a,b], Shiteng Zheng[c], Martin Treiber[d], Junfang Tian[a,b,*], Rui Jiang[c,*]

[a]*College of Management and Economics, Tianjin University, Tianjin, 300072, China*

[b]*Laboratory of Computation and Analytics of Complex Management Systems (CACMS), Tianjin University, Tianjin, 300072, China*

[c]*Key Laboratory of Transport Industry of Big Data Application Technologies for Comprehensive Transport, Ministry of Transport, Beijing Jiaotong University, Beijing 100044, China*

[d]*Institute for Transport and Economics, Dresden University of Technology, D-01062 Dresden, Germany*

*Corresponding author.

E-mail address: jftian@tju.edu.cn (JF. Tian), jiangrui@bjtu.edu.cn (R. Jiang).

# On the calibration of stochastic car following models

Recent experimental and empirical observations have demonstrated that stochasticity plays a critical role in car following (CF) dynamics. To reproduce the observations, quite a few stochastic CF models have been proposed. However, while calibrating the deterministic CF models is well investigated, studies on how to calibrate the stochastic models are lacking. Motivated by this fact, this paper aims to address this fundamental research gap. Firstly, the CF experiment under the same driving environment is conducted and analyzed. Based on the experimental results, we test two previous calibration methods, i.e., the method to minimize the Multiple Runs Mean (*MRMean*) error and the method of maximum likelihood estimation (*MLE*). Deficiencies of the two methods have been identified. Next, we propose a new method to minimize the Multiple Runs Minimum (*MRMin*) error. Calibration based on the experimental data and the synthetic data demonstrates that the new method outperforms the two previous methods. Furthermore, the mechanisms of different methods are explored from the perspective of error analysis. The analysis indicates that the new method can be regarded as a nested optimization model. The method separates the aleatoric errors caused by stochasticity from the epistemic error caused by parameters, and it is able to deal with the two kinds of errors effectively. Finally, we find that under the calibration framework of stochastic CF models, the calibrated parameter set using spacing as MoP may not always outperform that using velocity as MoP. These findings are expected to enhance the understanding of the role of stochasticity in CF dynamics where the new calibration framework for stochastic CF models is established.

## 1 Introduction

As one of the bases of traffic flow research, car following (CF) models are proposed to mimic the complex driving behavior interacting with vehicles and the driving environment by physical formulation from the longitudinal perspective. To evaluate and compare the ability of the CF models to characterize real traffic flow, the calibration and validation processes are indispensable. Calibration aims to find one set of model parameters with which the simulation outputs are best consistent with the observed data. Hence, calibration results refer to the best description of the model for the unique dataset.

To calibrate the deterministic CF models, e.g., the intelligent driver model (IDM, Treiber et al., 2000), the optimal velocity model (OVM, Bando et al., 1995), the Full Velocity Difference Model (FVDM, Jiang et al., 2001), and Newell's model (Newell, 2002), many remarkable achievements have been made under the deterministic calibration framework. As a systematic process consisting of several elements, formulation of the calibration (Sharma et al., 2019) is equivalent to the following optimization problem:

$$\text{minimize } GoF(MoP^{sim}(\beta), MoP^{obs})$$
$$\text{subject to } LB_\beta \leq \beta \leq UB_\beta \tag{1}$$

There are four core components in the calibration: the objective function, the dataset, the optimization algorithm (OA), and the CF model. The objective function is the combination of the measure of performance (MoP) and goodness of fitness (GoF). The controversy on the selection of MoP is not settled until Punzo and Montanino (2016) proved that the errors on spacing are cumulative on velocity and are more robust and therefore should be adopted as MoP. Meanwhile, a new comparison method of GoF based on Pareto efficiency (Punzo et al., 2021) is proposed which demonstrates that GoFs based on non-percentage errors are preferred over percentage-based ones. The above findings unified the foundation of the selection of objective function. In particular, recently, owing to the bountiful experimental and empirical datasets such as NGSIM, Naples Dataset (Punzo and Simonelli, 2005), Hefei Dataset (Jiang et al., 2015), and HighD dataset (Krajewski et al., 2018), the calibration research is greatly promoted. Furthermore, a wide variety of optimization algorithms are deployed, including the simultaneous perturbation stochastic approximation (SPSA) method (Lee and Ozbay, 2009), genetic algorithm (GA), OptQuest (Punzo et al., 2012), sequential quadratic programming (SQP) algorithm (Wang et al., 2010), DIviding RECTangles (DIRECT) algorithm (Li et al., 2016), Cross-Entropy Method (CEM) (Zhong et al., 2016), Levenberg-Marquardt algorithm (LM) (Treiber and Kesting, 2018), etc. According to the "No free lunch theorems for optimization", no algorithm can outperform others in all domains of calibration. Hence, the optimization algorithm should be specifically selected for the calibration scenario and problem. Ideally, the global optimum should be found which can be tested by running the calibration with different initial solutions.

In addition to the above components, calibration also involves many other aspects. To name a few, Treiber and Kesting (2013a) investigated the influence of data sample intervals and smoothing on calibration. Punzo et al. (2015) revealed the interaction mechanism of parameters, trajectories, and calibration outputs by applying global sensitivity analysis and proposed a simple model with fewer parameters. The completeness of the trajectories (i.e., combinations of driving regimes) in the data has also been proven to have an impact on the calibration results (Sharma et al., 2019) by using similar methods. For a comprehensive systematic review on calibration, one can refer to Punzo et al. (2021), in which all the core components are thoroughly explored and summarized under the calibration framework for deterministic CF models.

All these above findings are achieved in the deterministic calibration framework. However, recent CF experiments reveal that stochastic factors play a critical role in reproducing empirical CF dynamics. For instance, Jiang et al. (2014) demonstrated that the standard deviation of speed grows in a concave way along the observed CF platoon from the Hefei experiments. The concave pattern has also been confirmed by Tian et al. (2016a) by using the US 101 data which indicates it is a universal property (Zhou et al., 2017). In terms of microscopic mechanism, the concave growth pattern is due to the cumulative effect of stochastic factors (Jiang et al., 2018; Tian et al., 2019). Meanwhile, many stochastic CF models have been developed in recent years ( Laval et al., 2014; Tian et al., 2016b, 2019; Xu and Laval, 2020; Lee et al., 2019, 2021; Jiang et al., 2014; Treiber and Kesting, 2018). Unfortunately, stochasticity may add noises to the calibration process (Punzo et al., 2021). Thus, the previous calibration methods may be unsuitable. The lack of related research may impede the promotion and application of the stochastic CF models.

Motivated by the research gap, in this paper, we aim to propose a reasonable calibration framework for stochastic CF models. First, we review the previous calibration methodology for stochastic CF models to extract the current research gap. Then an additional dataset from our recent CF experiment is introduced and analyzed. The elements of calibration are presented. Moreover, we propose a new Multiple Runs Minimum (*MRMin*) method and compare it with two previous calibration methods based on the experimental trajectories. Next, the synthetic trajectories are generated by simulations through the CF models. The results of calibration on synthetic trajectories are highly consistent with the preset parameters, which demonstrates that the new method is numerically effective for stochastic model calibration. Furthermore, the new mechanism is explored from the perspective of error analysis, which explains why previous calibration methods lead to unreasonable results and the new method does not. Finally, we report the finding that under the new framework of stochastic CF model calibration, the calibrated parameter set using spacing as MoP may not always outperform that using velocity as MoP.

The contribution of this paper is threefold. (i) To establish a framework for calibrating stochastic CF models. (ii) To reveal the mechanism of different calibration methods from the perspective of error component analysis. (iii) To illustrate the selection of MoP.

The rest of the paper is organized as follows. Section 2 reviews the previous calibration methods for stochastic CF models and introduces the new experimental data and calibration settings. Section 3 investigates the performance of different calibration methods of the stochastic CF model. Section 4 presents the error analysis to explain the mechanism of the different calibration methods. In Section 5, we discuss the selection of MoP. Finally, the conclusion is summarized in Section 6.

## 2 Background and calibration settings

### 2.1 *Previous calibration methods*

Recently, a few stochastic models have been proposed, some of which have been calibrated but the process is still not unified. Laval et al. (2014) proposed a parsimonious stochastic CF model to describe the traffic oscillations and recommended the calibration method of performing a large number of runs to estimate the distribution of predicted trajectories. Xu and Laval (2019) advised to use Maximum Likelihood Estimation (*MLE*) for calibration. Shortly afterward, Xu and Laval (2020) estimated the parameters of the two-regime stochastic car-following models by *MLE*. Lee et al., (2019a) also adopted the *MLE* to calibrate the integrated deep learning and stochastic car-following model. Another estimation method is to use the mean simulation output of multiple runs for the objective function. In Lee et al. (2021), the mean of the simulated individual velocity among 100 runs is used to compare with the observed speed. Treiber and Kesting (2018) also selected the mean errors between the observed and simulated speed over 10 simulation runs with independent seeds as the objective function.

In addition to the calibration of microscopic trajectories of stochastic CF models, some calibration practices for macroscopic stochastic models and macroscopic patterns reproduced by stochastic CF models are conducted. To name a few, Ngoduy (2021) used the difference between the mean predicted and observed speed as the objective function to calibrate the stochastic higher-order continuum traffic models. Zheng et al. (2022) calibrated the models from the perspective of macroscopic patterns by comparing repeated simulated trajectories and the real ones aiming to reduce the impact of stochasticity on statistics. Apart from the above-mentioned study, several studies on the stochastic CF models do not specify the calibration settings. All these researches are summarized in Table. 1.

To summarize, two current calibration approaches for estimating the stochastic parameters of CF models are as follows:

- Maximum Likelihood Estimation (*MLE*)

The *MLE* is proposed by Hoogendoorn and Hoogendoorn (2010) which assumes the residual $e_k = \text{state}_k^{sim} - \text{state}_k^{obs}$ follows the zero-mean multivariate normal distribution with a covariance matrix $\Sigma$ and the residuals at different time steps $k$ are homoscedastic (constant $\Sigma$) and independent from each other. This leads to the following condition:

$$\begin{aligned} &\min z \\ &z = -L(\beta) \\ &L(\beta) = -\frac{K}{2}\ln(2\pi) - \frac{K}{2}\ln(\det(\Sigma)) - \sum_{k=1}^{K} e_k^T \Sigma^{-1} e_k \\ &\textit{subject to:} \beta_{\min} \leq \beta \leq \beta_{\max} \end{aligned} \quad (2)$$

The covariance is estimated by $\Sigma = \frac{1}{K}\sum_{k=1}^{K} e_k e_k^T$.

2  Table. 1. Review of the calibration settings for stochastic CF models.

| Microscopic Calibration | Data | MoP | GoF | Model | OA | Estimation Method |
|---|---|---|---|---|---|---|
| Treiber and Kesting (2018) | 25-car-platoon experiment | $\sigma_v$ | SSE | IDM with stochastic acceleration noise and action points | Levenberg-Marquardt | Mean simulated value of 10 repeated runs |
| Ngoduy et al. (2019) | NGSIM | V | *RMSE* | OVM with stochastic acceleration noise by Cox-Ingersoll-Ross process | Genetic algorithm | Mean simulated value of 100 runs |
| Lee et al. (2019) | NGSIM | V | Likelihood Function | Same as Ngoduy et al., (2019) | - | *MLE* |
| Xu and Laval (2019) | 25-car-platoon experiment | - | - | Two-Regime Stochastic Car-Following Model | - | *MLE* recommended |
| Xu and Laval (2020) | 7-car-platoon experiment & 25-car-platoon experiment | S | Likelihood Function | Two-Regime Stochastic Car-Following Model | - | *MLE* |
| Lee et al. (2021) | Circular road experiment | V | *RMSE* | Stochastic behavior model of personal mobility | Genetic algorithm | Mean simulated value of 100 runs |
| Ngoduy (2021) | NGSIM | V | Index PI | Stochastic higher-order continuum traffic models | Genetic algorithm | Mean simulated value |
| Zheng et al., (2022) | Circular road experiment | $\sigma_v$ and macroscopic variables | Mixed macroscopic relative errors | 2D-IDM and SSAM | Genetic algorithm | Mean simulated value of M runs |

- Multiple Runs Mean Value Method (*MRMean* for short):

In *MRMean*, the mean value of the GoF in multiple replicated simulations is used as the optimization objective and the formula is as follows:

$$\min z$$

$$z = \frac{\sum_{i=1}^{N} GoF(MoP^{obs}, MoP_i^{sim})}{N} \quad \text{or} \quad GoF(MoP_i^{obs}, \frac{\sum_{i=1}^{N} MoP_i^{sim}}{N}) \quad (3)$$

$$MoP_i^{sim} = F(\beta)$$

$$\text{subject to:} \beta_{\min} \leq \beta \leq \beta_{\max}$$

where $MoP^{obs}$ is the observed *MoP*, $MoP_i^{sim}$ is the simulated MoP in the *i*th run. $N$ is the number of repeated simulations. When the spacing is adopted as MoP and the *RMSE* is adopted as GoF, the objective function will be:

$$z = \frac{\sum_{i=1}^{N} RMSE_s^i}{N} = \frac{\sum_{i=1}^{N} \sqrt{\frac{1}{K} \sum_{k=1}^{K} (s_{i,k}^{sim} - s_k^{obs})^2}}{N} \quad (4)$$

where $s_k^{obs}$ is the spacing between the car and the preceding car in the *k*th time step, obtained by observation, $s_{i,k}^{sim}$ is the spacing in the *k*th time step obtained by simulation in the *i*th run. $k \in [0, K]$. $\Delta t$ is the time step and the total time is $K\Delta t$.

## 2.2 Experimental data

The essential difference between the stochastic models and the deterministic models is that the model output varies even under the same input. Inspired by this fact and to investigate the property of stochasticity of driving behavior, one CF experiment under the same driving environment is carried out.

The experiment was performed on October 12, 2021, on a straight road of about 1.5 km in length in the Traffic Test Site of the Ministry of Transport, China. In the experiment, high-precision global navigation systems (precision $\pm 1$ m for location and $\pm 1$km/h for velocity) have been installed on the cars to record their locations and velocities every 0.1 s. One autonomous car (AV) and one human-driven car (HV) are involved. A total of nine runs are conducted in this experiment under the same circumstance. Initially, the two cars are stopped bumper-to-bumper. In each run, the AV moves as the leading car and follows the same designed trajectory with the same control parameters. The HV is driven by the same driver to follow the AV. Hence the external driving scenario of the HV in the nine runs is almost the same. Notably, the pre-runs are important for drivers to get more experienced with the test protocol, especially for those complex experiments of long periods. However, the setup of this experiment was relatively simple so pre-runs experiments are not needed. The trajectories of both cars are collected by high-precision GPS. There is no obvious adaptive behavior over the nine runs which can be seen from the velocity and spacing profile in Fig. 1 and Fig. 2. Meanwhile, all the trajectories contain four driving regimes: two times for acceleration (**A**), three times for deceleration (**D**), four times for following (**F**), and one time for standstill (**S**) so that the trajectory completeness of this experiment belongs to the **ADFS** (Sharma et al., 2019) which is the most common one in the reconstructed NGSIM I-80 data (Sharma et al., 2018).

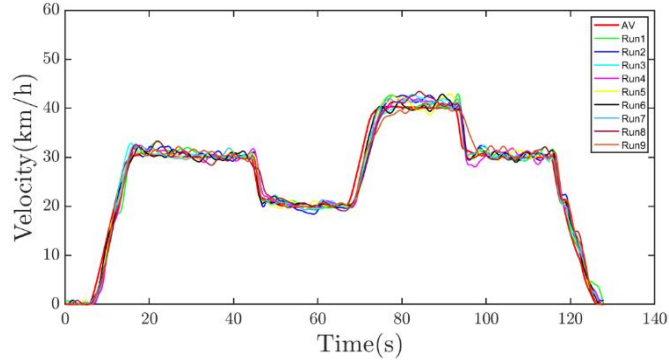

Fig. 1. The velocity profile of the experimental data. The red lines represent the leading AV and the other colors represent the following HV over the nine runs.

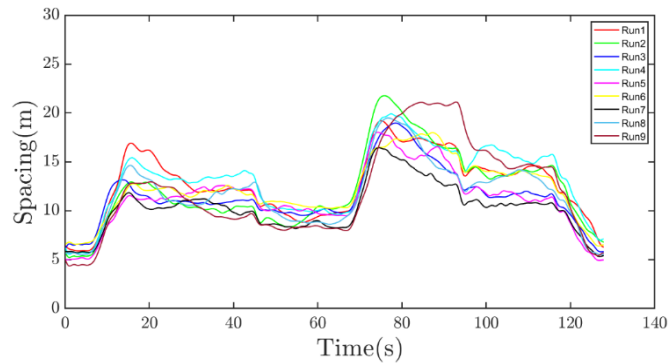

Fig. 2. The spacing profile between AV and HV over the nine runs.

To further analyze the stochasticity of car-following behavior, we use the paired t-test[1] at 95% significance to compare the difference among different runs. One can see from Table. 2 and Table. 3 that none of the velocity sequences for the nine trajectories differed significantly. In contrast, most of the trajectories are significantly different in spacing. The result indicates that the driver has more tendency to keep a similar velocity rather than similar spacing in the same driving environment.

---

[1] The spacing or velocity sequence is non-stationary time series. The paired t-test can focus more on the difference between pairs of data at each time step which effectively extracts the sequence pattern. While non-parametric methods (such as the KS test) calculate the similarity of the cumulative distribution of the whole sequence, they may be incapable of extracting the time-related information.

In real traffic flow, we can only acquire one trajectory. This raises the following two important questions:

Q1. How to calibrate the driving behavior with stochasticity against one single trajectory?

Q2. Whether the different experimental trajectories can be characterized by one set of parameters of the stochastic model?

Table. 2. The significance test of the spacing sequence of nine trajectories by using paired t-test at 95% confidence level. 0 and 1 represent no significant difference and significant difference, respectively.

| Spacing | 1 | 2 | 3 | 4 | 5 | 6 | 7 | 8 | 9 |
|---|---|---|---|---|---|---|---|---|---|
| 1 | 0 | 1 | 1 | 1 | 1 | 1 | 1 | 1 | 1 |
| 2 |   | 0 | 1 | 1 | 1 | 0 | 1 | 1 | 0 |
| 3 |   |   | 0 | 1 | 0 | 1 | 1 | 1 | 1 |
| 4 |   |   |   | 0 | 1 | 1 | 1 | 1 | 1 |
| 5 |   |   |   |   | 0 | 1 | 1 | 1 | 1 |
| 6 |   |   |   |   |   | 0 | 1 | 1 | 0 |
| 7 |   |   |   |   |   |   | 0 | 1 | 1 |
| 8 |   |   |   |   |   |   |   | 0 | 0 |
| 9 |   |   |   |   |   |   |   |   | 0 |

Table. 3. The significance test of the velocity sequence of nine trajectories by using paired t-test at 95% confidence level. 0 and 1 represent no significant difference and significant difference, respectively.

| Velocity | 1 | 2 | 3 | 4 | 5 | 6 | 7 | 8 | 9 |
|---|---|---|---|---|---|---|---|---|---|
| 1 | 0 | 0 | 0 | 0 | 0 | 0 | 0 | 0 | 0 |
| 2 |   | 0 | 0 | 0 | 0 | 0 | 0 | 0 | 0 |
| 3 |   |   | 0 | 0 | 0 | 0 | 0 | 0 | 0 |
| 4 |   |   |   | 0 | 0 | 0 | 0 | 0 | 0 |
| 5 |   |   |   |   | 0 | 0 | 0 | 0 | 0 |
| 6 |   |   |   |   |   | 0 | 0 | 0 | 0 |
| 7 |   |   |   |   |   |   | 0 | 0 | 0 |
| 8 |   |   |   |   |   |   |   | 0 | 0 |
| 9 |   |   |   |   |   |   |   |   | 0 |

### 2.3 Optimization algorithm

As another element of calibration, the optimization algorithm is of critical role. In recent years, several improved heuristic algorithms have been proposed. Ciuffo and Punzo (2014) found that the GA outperforms the others globally by comparing different OAs and it is also the most commonly used one in calibrating stochastic CF models (see Table. 1). Therefore, we applied GA as the optimization algorithm in this paper by using the MATLAB toolbox.

## 2.4 MoP and GoF

In sections 3.1 and 3.2, to test different calibration methods, the *RMSE* is selected as the GoF, since it can avoid situations where the error is magnified by an observed MoP of zero (e.g., when the observed velocity is zero as the denominator of the formula of *RMSPE*). The spacing is selected as the MoP, as suggested by Punzo and Montanino (2016).

## 2.5 Model and its integration scheme

2D-IDM is selected for the calibration to ensure the adaptability and generalization of the results:

- 2D-IDM inherits the excellent properties of IDM in which each parameter has a clear physical meaning.
- The stochastic items in 2D-IDM are in a neat form and easy to understand. By adopting the time-varying time headway, it can capture the stochasticity of driving behavior.
- 2D-IDM has been shown, as a stochastic CF model, to be capable of reproducing the concave growth pattern, i.e, the traffic oscillation grows in a concave way among the CF platoon.
- It can span a 2D region in the velocity-spacing plane so that the dynamic relation of velocity and spacing can be described.

The mathematical formulation of the 2D-IDM is as follows (Jiang et al., 2014):

$$a_n(t) = a\left[1 - \left(\frac{v_n(t)}{v_{max}}\right)^4 - \left(\frac{s_n^*(t)}{\Delta x_n(t)}\right)^2\right] \quad (5)$$

$$s_n^*(t) = s_0 + \max\left(T_n(t)v_n(t) + \frac{v_n(t)\Delta v_n(t)}{2\sqrt{ab}}, 0\right) \quad (6)$$

$$T_n(t+\Delta t) = \begin{cases} T_1 + r(t)(T_2 - T_1) & r_1(t) < p\Delta t \\ T_n(t) & r_1(t) \geq p\Delta t \end{cases} \quad (7)$$

where $a_n$, $v_n$, $T_n$ and $s_n^*$ is the acceleration, velocity, desired time headway, and desired spacing of the *n*th car, respectively. $r(t)$ and $r_1(t)$ are two uniformly distributed pseudorandom numbers in [0,1]. $\Delta v_n(t) = v_n(t) - v_{n-1}(t)$ is the velocity difference between the following car *n* and the preceding car $n-1$. The definition and calibration boundaries of parameters of 2D-IDM are shown in Table. 4. Note that $T_2 = T_1 + \Delta T$.

Table. 4. Parameters definition and calibration boundaries of experimental trajectories for 2D-IDM.

| Parameters | Definition | Unit | Calibration Boundaries |
|---|---|---|---|
| $v_{max}$ | Maximum Velocity | (km/h) | [40,60] |
| $a$ | Maximum Acceleration | (m/s$^2$) | [0.5,3] |
| $b$ | Maximum Deceleration | (m/s$^2$) | [0.5,5] |
| $s_0$ | Minimum spacing | (m) | [0.5,5] |
| $T_1$ | Minimum desired time headway | (s) | [0.1,1] |
| $\Delta T$ | The changeable interval of desired time headway | (s) | [0,1.3] |
| $p$ | randomization change rate | (s$^{-1}$) | [0,1] |

Ballistic updating is adopted as the model integration scheme, which has been proved to consistently outperform the Euler method (Treiber and Kanagaraj, 2015):

$$\begin{cases} v(t+\Delta t) = v(t) + a(t)\Delta t \\ x(t+\Delta t) = x(t) + \dfrac{v(t)+v(t+\Delta t)}{2}\Delta t \end{cases} \quad (8)$$

## 3 The stochastic CF model Calibration

### 3.1 The Multiple Runs Minimum Method

Apart from the two calibration methods introduced in section 2.1, a new calibration method is proposed, which has a different mechanism from the view of driving stochasticity. As we mentioned in section 2.2, the trajectories of multiple runs can be regarded as different realizations of the stochastic CF model. Hence, the Multiple Runs Minimum Method (*MRMin* for short) takes the minimum value of the GoF (the best realization of the CF model) in multiple replicated simulations as the optimization objective:

$$\begin{aligned} &\min z \\ &z = \min_{i \in [1,\ldots,N]} \{GoF(MoP^{obs}, MoP_i^{sim})\} \\ &MoP^{sim} = F(\beta) \\ &subject\ to: \beta_{min} \leq \beta \leq \beta_{max} \end{aligned} \quad (9)$$

When *RMSE* is adopted as GoF and spacing is adopted as MoP, one has:

$$\min z = \min_{i \in [1,\ldots,N]}\{RMSE_s^i\} = \min_{i \in [1,\ldots,N]} \sqrt{\dfrac{1}{K}\sum_{k=1}^{K}(s_{i,k}^{sim} - s_k^{obs})^2} \quad (10)$$

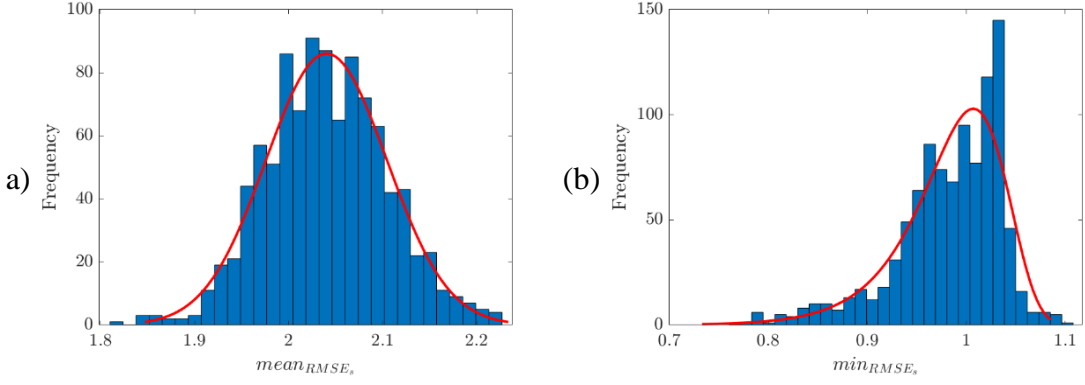

Fig. 3. The simulated distribution of objective function (a) *MRMean* and (b) *MRMin* on spacing when *N*=1000. The sample number is 1000. The red line is a fitting curve, which shows the normal distribution in panel (a) and the Gumbel distribution in panel (b).

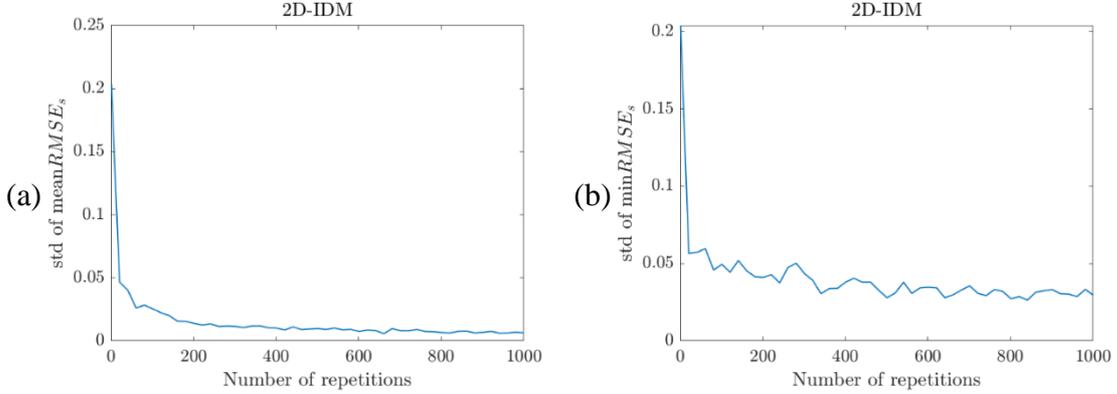

Fig. 4. The impact of *N* on the stability of objective function with (a) *MRMean* and (b) *MRMin* on spacing. The sample number is 1000.

The calibration settings have been clarified in Section 2. Note that in Table. 1, the number of repeated simulations was usually set arbitrarily. If we assume the *RMSE* on spacing follows $N(\mu,\sigma^2)$, according to the central limit theorem, the *MRMean* of *RMSE* on spacing can be regarded as a random variable following $N(\mu,\frac{\sigma^2}{N})$, see Fig.3(a). Hence, for *MRMean*, the standard deviation of *N*=*n* is $\frac{1}{\sqrt{n}}$ times the standard deviation of *N*=1. Therefore, only exponential growth of *N* can bring precision benefits for *MRMean*, see Fig. 4. The *MRMin* can be regarded as extreme values. In Fig. 3(b), one can see that it is well approximated by extreme value distribution (the Gumbel distribution). We calculate the standard deviations of the objective function of *MRMin* under different values of *N* in Fig. 4. It can be seen that the larger *N* is, the more stable the objective function is. To summarize, the increase of *N* brings small stability benefits when *N* is large and the calibration computation time will increase greatly. Therefore, *N* is uniformly set to 200 in this paper.

*3.2 Calibration methods test using experimental data*

For each trajectory in the nine runs, a set of parameters is calibrated to examine the validity of the calibration approach and the difference between the calibrated parameter sets. Given that each trajectory is collected under the same condition, the value of calibrated parameter sets in the nine runs should be:

C1 Concentrated. Given that the nine trajectories are generated by one driver under the same circumstance, the parameter sets of the nine trajectories should be close to each other.

C2 Valid with stochasticity. (i) the stochastic parameter ΔT should not be extremely small. Otherwise, the model would tend to be deterministic; (ii) and the parameter p should not be too large to be consistent with reality.

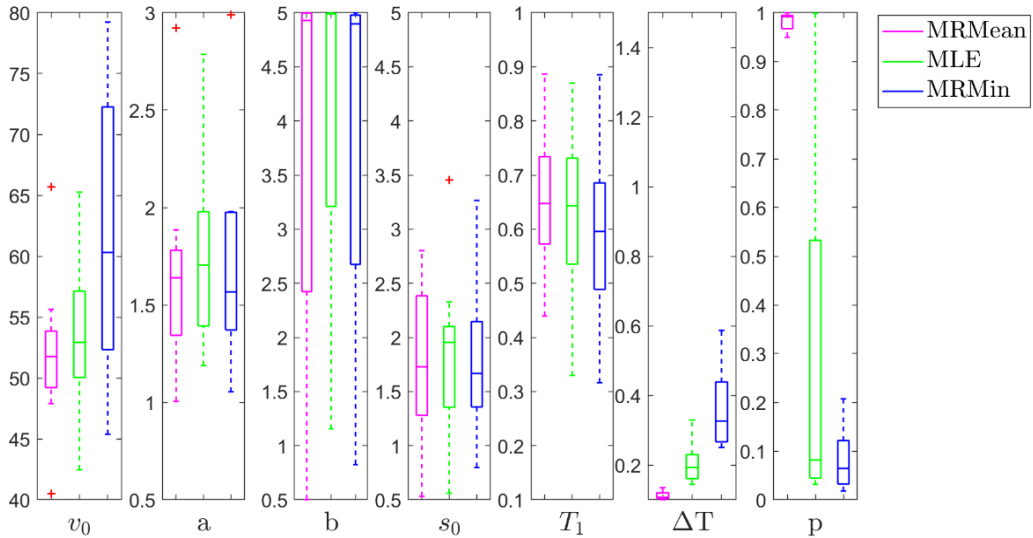

Fig. 5. Boxplot of calibration results for experimental trajectories by the method of *MRMean*, *MLE* , and *MRMin*. The calibration boundaries are presented in Table. 4.

The calibration results of the nine experimental trajectories are shown as boxplots in Fig. 5. It can be seen that when applying *MRMean*, the distribution of the parameter *p* is concentrated and close to the upper boundary, and the parameter $\Delta T$ is small. This indicates that the calibrated 2D-IDM tends to be deterministic. It can be observed more explicitly in the simulated trajectories in Fig. 6 (a). The 5%-95% prediction (simulated) interval under one set of parameters and the profile of prediction data is approximately one curve rather than a band. From the view of driving behavior, the *MRMean* is represented to minimize the averaged driving behavior which may likely neutralize or even eliminate the driving behavior stochasticity and is contrary to the physical meaning of stochasticity in driving behavior. Similarly, Punzo and Montanino (2020) also pointed out that using the averages of many scenario simulations to represent the traffic system makes no sense, which is an incomplete or artefactual system representation.

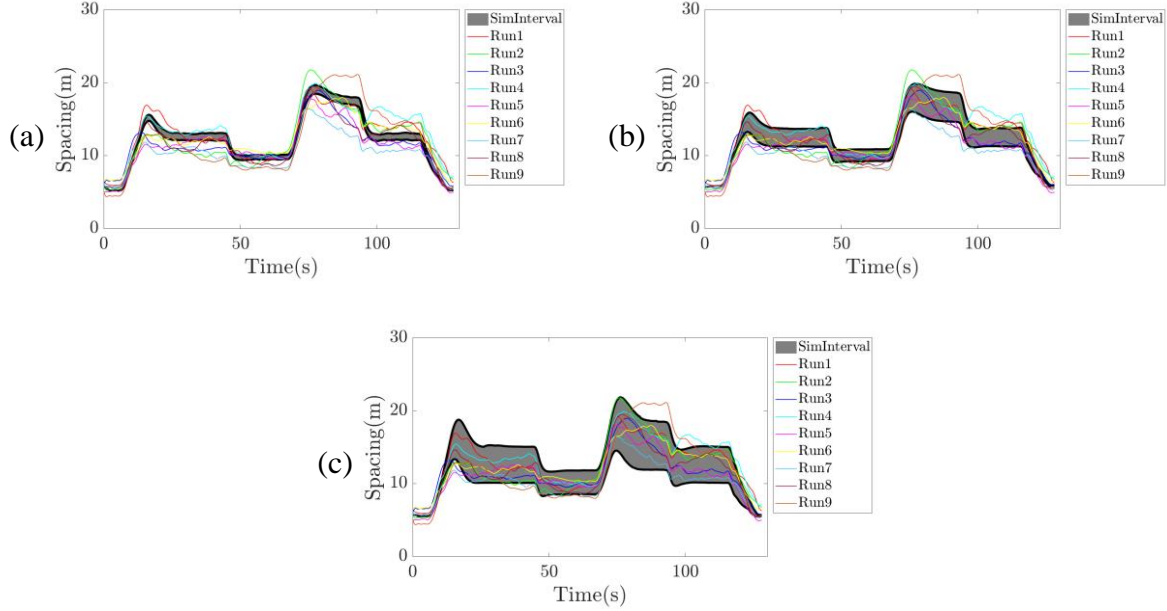

Fig. 6. Simulated spacing profile of 5-95% prediction interval of 2D-IDM by using one sample set of parameters from the calibrated results of (a)*MRMean* (b)*MLE* (c)*MRMin*.

When it comes to *MLE*, simulation results indicate *MLE* can keep the stochasticity of the model. Fig. 6 (b) demonstrates that the prediction band of the parameters calibrated with *MLE* can partially cover the nine real trajectories. Finally, Fig. 6 (c) shows that the prediction band of the parameters calibrated with *MRMin* can cover the nine real trajectories better. This demonstrates that the *MRMin* performs better than the *MLE*.

Even if the stochastic parameters are successfully calibrated by *MRMin* in plausible values and consistent with criteria C2, it does not necessarily mean that the estimation of stochastic parameters is reasonable because we cannot figure out what the actual value of parameters should be. Hence, there are still two problems to be addressed.

M1. The numerical correctness of *MRMin*. The synthetic trajectories can be used for calibration to verify the consistency between the calibrated value and preset/ground truth value.

M2. The calibrated parameters may fall into the trap of overfitting and perform poorly on the training datasets. The only way out of this trap is validation.

Hence, in sections 3.3 and 3.4, the numerical correctness and efficiency of the new method are examined.

### 3.3 *Numerical correctness tests on calibration methods by using synthetic data*

To evaluate the validity and correctness of the *MRMin*, we synthesized 30 trajectories by using the experimental AV trajectories as leading car trajectories and generated the synthetic trajectories of the following car through simulations of 2D-IDM. Since it aims only to check the consistency between the parameter sets calibrated by different calibration methods and the preset parameters of the synthetic trajectory (mainly for the stochastic parameters), only the three parameters related to stochasticity are calibrated. Other parameters are given. The preset value and calibration boundaries of parameters are presented in Table. 5.

Table. 5. The parameters and calibration boundaries of synthetic trajectories for 2D-IDM

| parameters | $v_{max}$ | $a$ | $b$ | $s_0$ | $T_1$ | $\Delta T$ | $p$ |
|---|---|---|---|---|---|---|---|
| Preset parameters | 50 | 1.5 | 2.5 | 2 | 0.6 | 0.5 | 0.1 |
| calibration boundaries | | | | | [0.1,1] | [0.1,1.5] | [0,1] |

Fig. 7 shows that the three stochastic parameters calibrated by *MRMean* lead to the degradation of stochastic CF models, which is similar to the results in Fig. 5. As for *MLE*, roughly speaking, the stochastic parameter $p$ is calibrated pretty well. However, there are still two erratic outliers. Moreover, $\Delta T$ calibrated by *MLE* deviates significantly from the real value, which indicates the inaccuracy of *MLE*[2]. In contrast, the three parameters are all calibrated pretty well by *MRMin*.

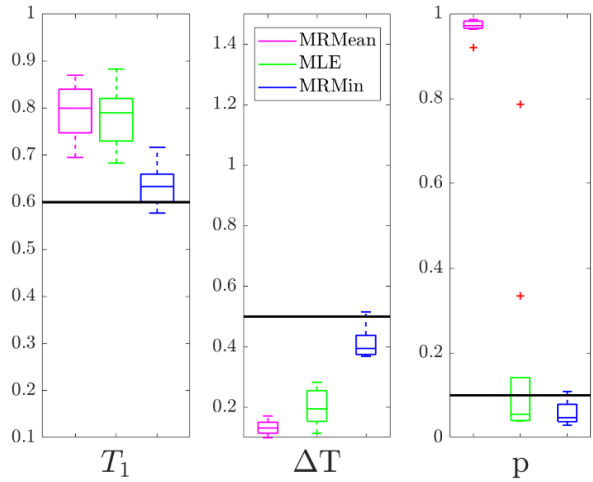

Fig. 7. Boxplot of calibration results for synthetic trajectories of 2D-IDM by the method of *MRMean*, *MLE*, and *MRMin*. The black line represents the preset parameters.

In the deterministic CF models, the synthetic trajectories can always be calibrated with zero error. In contrast, there are nonzero errors in all three methods, as shown in Fig. 7. Nevertheless, the preset parameters can be calibrated effectively by the *MRMin* method. The calibrated results are all concentrated around the preset parameters. To further illustrate the property and validity of the objective function, the contour plot is shown in Fig. 8. For *MRMin*, we can observe a relatively distinctive global minimum, which is almost equal to the preset parameters, see Fig. 8 (a1) and (a2). For *MRMean*, no matter how many times we re-sample the contour plots, it always shows a smooth and nearly the same shape, see Fig. 8 (b1) and (b2). The global optimum is achieved when $p=1$, which significantly deviates from the preset value. For *MLE*, the objective function is quite unstable and the shape of the contour plot is irregular. Therefore, it is incapable of finding a stable global optimum near the preset values, see Fig. 8(c1) and (c2). All these indicate that only the objective function of *MRMin* is effective to identify a global optimum near the preset parameters.

---

[2] This may be because the assumption of normal distribution of errors in *MLE* is not always suitable when calibrating the stochastic CF model. Thus, the estimation of the covariance matrix is inaccurate.

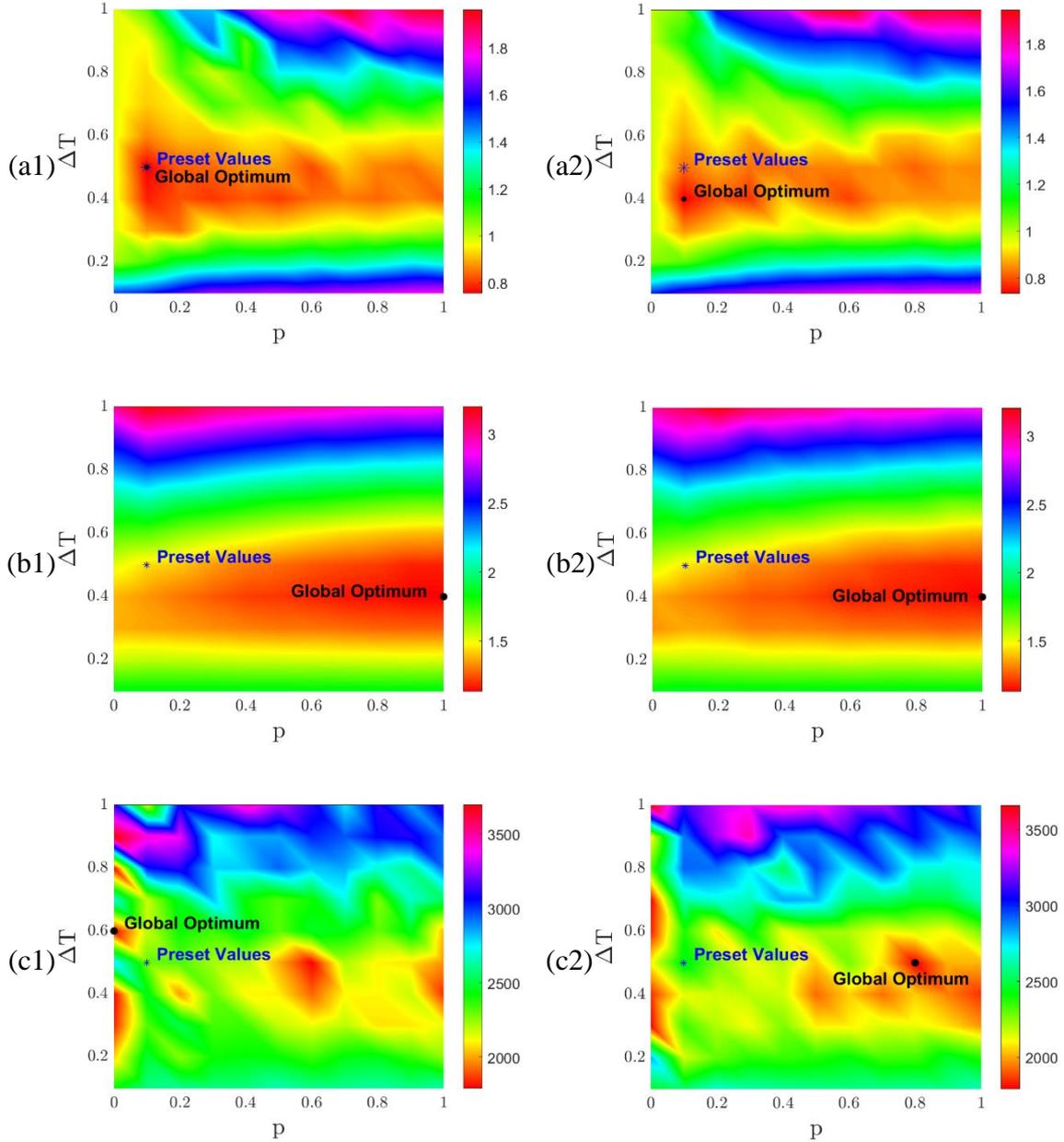

Fig. 8. Contour plots of objective functions of (a) *MRMin* (b)*MRMean* (c) *MLE* against $\Delta T$ and $p$. The blue star represents the preset value of $\Delta T$ and $p$. The black point represents the Global Optimum of the corresponding objective function. The other parameters except for $\Delta T$ and $p$ are set as the preset values. For each method, we show two contour plots to show the variation of profile under different random seeds due to the stochasticity.

## 3.4 Validation of the calibration results on experimental data

Section 3.3 has demonstrated that the *MRMin* is effective for the calibration of stochastic CF models, which answers question Q1 in section 2.2. This subsection investigates question Q2, i.e., Whether the different experimental trajectories can be characterized by one set of parameters of the stochastic model. To this end, we apply the two-level validation method proposed by Punzo and Montanino (2020) and select the best realization among *N* runs with the parameters calibrated by *MRMin*. As an example, Fig. 9 demonstrates that the best realization among the simulated trajectories of the 2D-IDM can well capture Run 1. Moreover, as shown in Appendix A, this is true for all the nine experimental trajectories for the same parameter set (and usually different simulated realization). This gives evidence for the hypothesis that the dynamics of all nine experimental runs are equivalent, i.e., all intra-driver (intra-run and inter-run) variations are captured by the random terms of the model.

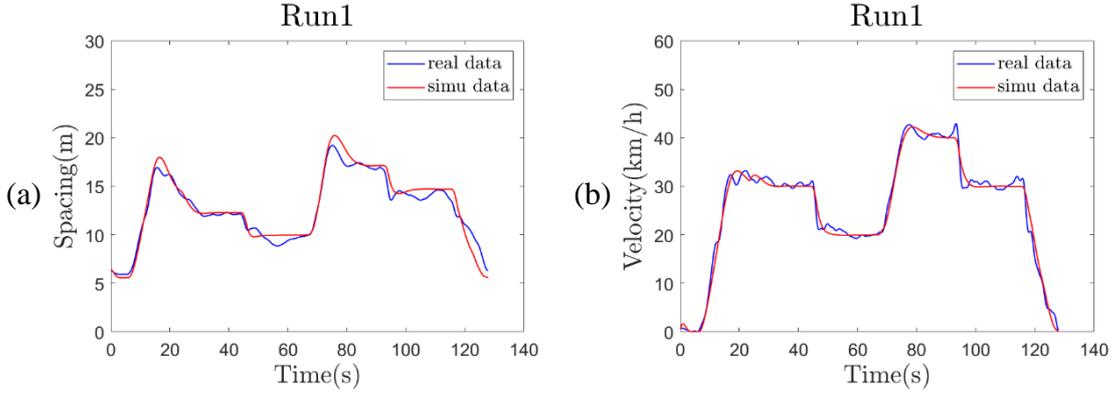

Fig. 9. The best-simulated realization of 2D-IDM among *N* runs with one set of parameters calibrated by *MRMin*.

We define the errors under the best realization among the model outputs of multiple runs as the validation error, which can be formulated as

$$ValiError\left(\beta^*, Traj_{obs}\right) = \min_{i \in [1,...,N]} RMSE\left(MoP_{Traj_{obs}}, MoP_i^{sim}\left(\beta^*\right)\right) \quad (11)$$

where $\beta^*$ is the calibrated parameter set, $Traj_{obs}$ is the observed (experimental or empirical) trajectory used for validation. In accordance with the calibration setup in section 3.2, we choose spacing as the MoP of validation. Table. 6 shows the validation errors of the *MRMin* method. The validation error for *MRMean* and *MLE* can be found in Appendix A. Table. 7 compares the three calibration methods. One can see that the *MRMin* outperforms *MLE* and *MRMean*. Meanwhile, the standard deviation of the validation errors of *MRMin* is the smallest, which further supports the hypothesis of Q2, i.e., no matter which trajectory is used for calibration, the parameter sets calibrated with *MRMin* can always capture the dynamic features of driving behavior. The results also demonstrate that there is no overfitting, which addresses the concern of M2.

Table. 6. The validation errors of 2D-IDM (Units: m) calculated by *MRMin* with parameter sets of different trajectories calibrated by *MRMin*. The column of "mean" represents the averaged validation performance of the parameter set calibrated with one trajectory on all the trajectories.

| Validation Error | | Validation errors calculated on trajectory No.* (Unit: m) | | | | | | | | | Mean | STD |
|---|---|---|---|---|---|---|---|---|---|---|---|---|
| | | 1 | 2 | 3 | 4 | 5 | 6 | 7 | 8 | 9 | | |
| Calibrated parameter set of trajectory No.* | 1 | 0.681 | 1.719 | 1.444 | 1.112 | 1.469 | 1.05 | 1.629 | 1.151 | 2.237 | 1.388 | 0.453 |
| | 2 | 1.022 | 0.883 | 0.944 | 1.03 | 1.039 | 0.945 | 0.841 | 0.884 | 1.859 | 1.05 | 0.312 |
| | 3 | 1.209 | 1.323 | 0.55 | 1.499 | 0.929 | 0.744 | 0.972 | 1.057 | 2.05 | 1.148 | 0.444 |
| | 4 | 0.956 | 1.529 | 1.53 | 1.017 | 1.666 | 0.95 | 2.29 | 1.224 | 2.185 | 1.483 | 0.503 |
| | 5 | 1.424 | 1.548 | 1.014 | 1.855 | 0.584 | 0.89 | 0.775 | 1.128 | 2.268 | 1.276 | 0.546 |
| | 6 | 1.147 | 1.699 | 1.18 | 1.287 | 0.908 | 0.441 | 1.286 | 1.221 | 2.079 | 1.25 | 0.458 |
| | 7 | 2.441 | 2.455 | 1.151 | 2.922 | 1.076 | 1.944 | 0.513 | 1.712 | 3.074 | 1.921 | 0.882 |
| | 8 | 0.958 | 1.396 | 1.109 | 1.383 | 0.926 | 1 | 1.107 | 0.712 | 2.242 | 1.204 | 0.445 |
| | 9 | 1.242 | 1.486 | 1.421 | 1.545 | 0.992 | 1.093 | 1.05 | 1.3 | 0.891 | 1.224 | 0.232 |
| Mean value of All | | | | | | | | | | | **1.327** | **0.475** |

Table. 7. The validation errors comparison of 2D-IDM (Units: m) calculated by *MRMin* with parameter sets of different trajectories calibrated by three calibration methods.

| Methods | Validation errors calculated on trajectory No.* | | | | | | | | | Mean of all | STD of all |
|---|---|---|---|---|---|---|---|---|---|---|---|
| | 1 | 2 | 3 | 4 | 5 | 6 | 7 | 8 | 9 | | |
| *MRMean* | 1.733 | 1.507 | 1.537 | 1.904 | 1.522 | 1.523 | 2.128 | 1.479 | 2.368 | 1.745 | 0.588 |
| *MLE* | 1.511 | 1.092 | 1.444 | 1.505 | 1.487 | 1.343 | 2.054 | 1.347 | 1.744 | 1.503 | 0.514 |
| *MRMin* | 1.388 | 1.05 | 1.148 | 1.483 | 1.276 | 1.25 | 1.921 | 1.204 | 1.224 | 1.327 | 0.475 |

## 4  Errors analysis

Table. 8. Variables and related notations.

| Notations | Explanation |
|---|---|
| $\Delta t$ | Time step |
| $K$ | Total time step |
| $K\Delta t$ | Total simulation time |
| $r(t), r_1(t)$ | Uniformly distributed pseudorandom numbers in [0,1] |
| $RS_i$ | The random seed in the $i^{th}$ repeated simulation |
| $e_k^{MoP}$ or $e_k^{MoP}(\beta)$ | Total epistemic error in the $k^{th}$ time step of any simulation |
| $\varepsilon_k^{MoP}$ or $\varepsilon_k^{MoP}(\beta)$ | Parameter epistemic error in the $k^{th}$ time step of any simulation |
| $\xi_k^{MoP}$ or $\xi_k^{MoP}(\beta)$ | Random variable: Aleatoric error in the $k^{th}$ time step |
| $\xi_k^{MoP}(RS_i)$ or $\xi_k^{MoP}(\beta, RS_i)$ | Random sample: Aleatoric error in the $k^{th}$ time step of the $i^{th}$ repeated simulation |
| $v_k^{sim}(\beta, RS_i)$ | Simulated velocity in the $k^{th}$ time step of the $i^{th}$ repeated simulation |
| $v_k^{obs}$ | Observed velocity in the $k^{th}$ time step |
| $x_k^{sim}(\beta, RS_i)$ | Simulated position in the $k^{th}$ time step of the $i^{th}$ repeated simulation |
| $x_k^{obs}$ | Observed position in the $k^{th}$ time step |
| $e = (e_1^v, e_2^v, \cdots, e_K^v)^T$ | Vector of the total epistemic error on the velocity of any simulation |
| $\xi(RS_i) = (\xi_1^v(RS_i), \xi_2^v(RS_i), \cdots, \xi_K^v(RS_i))^T$ | Vector of the aleatoric error on the velocity of the $i^{th}$ repeated simulation |
| $\sigma = (\sigma_1, \sigma_2, \cdots, \sigma_K)^T$ | Vector of the standard deviations of $\xi_k^{MoP}(\beta)$ |
| $\beta^*$ | The optimal/calibrated solution of the parameters set |
| $RS^*$ | The optimal/calibrated solution of random seed |

### *4.1  The components of errors*

Given a fixed leader trajectory and fixed initial speed and position of the follower, the follower's trajectory simulated by a deterministic model with a given parameter set is fixed. So is the calibration error. However, this is different for stochastic CF models. The total calibration errors of stochastic CF models consist of two components: epistemic error component $e$ and aleatoric component $\xi$. Both components are the function of the model, the parameters, and the selected MoP.

When conducting a simulation of a given stochastic CF model, the epistemic error component $e_k^{MoP} = e_k^{MoP}(\beta)$ includes the model epistemic uncertainty (the irreducible model performance gap) and the parameter epistemic error $\varepsilon_k^{MoP}(\beta)$ (the reducible gap between the present parameters and the optimal parameters). $e_k^{MoP}$ is independent of the random seed. No matter how many repetitions of simulations are conducted, the value of $e_k^{MoP}$ is always fixed under a given set of parameters. The aleatoric error component $\xi_k^{MoP} = \xi_k^{MoP}(\beta)$ contains the aleatoric uncertainty (the volatility due to model stochastic terms). $\xi_k^{MoP}$ is assumed as a centralized random variable, i.e., the expectation is zero and the variance is $\sigma^2(\beta, MoP)$. For the $i$th repeated simulation, $\xi_k^{MoP}(\beta, RS_i)$ is a random sample drawn by the random seed from its empirical distribution. If $RS_i$ is settled, i.e., the sequence of $r(t)$ and $r_1(t)$ is fixed, the output will be fixed. Hence, the parameter set determines the empirical distribution of $\xi^{MoP}$ while $RS_i$ (RS of the $i^{\text{th}}$ simulation) determines its value $\xi^{MoP}(\beta, RS_i)$ in one specific simulation realization. To summarize, in the process of solving the calibration problem, the simulated velocity of the $k^{\text{th}}$ time step in the $i^{\text{th}}$ repeated simulation can be written as follows:

$$v_k^{sim}(\beta, RS_i) = v_k^{obs} + \underbrace{e_k^v(\beta)}_{\text{Total Epistemic Error}} + \underbrace{\xi_k^v(\beta, RS_i)}_{\text{Aleatoric Error}}$$

$$= \underbrace{v_k^{opt}(\beta^*)}_{\text{Optimal Velocity}} + \underbrace{\varepsilon_k(\beta)}_{\text{Parameter Epistemic Error}} + \xi_k^v(\beta, RS_i) \quad (12)$$

In fact, the component we want to minimize is the epistemic error. To find the impact of error components on the objective function of calibration, we conduct the error propagation formulation (see, e.g., Punzo and Montanino, 2016). For the sake of simplicity, $\beta$ is omitted in the formula with ballistic update rules:

$$\begin{cases} v_1^{sim}(RS_i) = v_1^{obs} + e_1^v + \xi_1^v(RS_i) \\ x_1^{sim}(RS_i) = x_0^{sim}(RS_i) + \dfrac{v_1^{sim}(RS_i) + v_0^{sim}(RS_i)}{2}\Delta t \\ \quad = x_0^{obs} + \dfrac{v_1^{obs} + v_0^{obs}}{2}\Delta t + \dfrac{e_1^v + \xi_1^v(RS_i)}{2}\Delta t = x_1^{obs} + \dfrac{e_1^v + \xi_1^v(RS_i)}{2}\Delta t \end{cases}$$

$$\begin{cases} v_2^{sim}(RS_i) = v_2^{obs} + e_2^v + \xi_2^v(RS_i) \\ x_2^{sim}(RS_i) = x_1^{sim}(RS_i) + \dfrac{v_2^{sim}(RS_i) + v_1^{sim}(RS_i)}{2}\Delta t \\ \quad = x_1^{obs} + \dfrac{e_1^v + \xi_1^v(RS_i)}{2}\Delta t + \dfrac{v_2^{obs} + v_1^{obs}}{2}\Delta t + \dfrac{e_2^v + \xi_2^v(RS_i)}{2}\Delta t + \dfrac{e_1^v + \xi_1^v(RS_i)}{2}\Delta t \\ \quad = x_2^{obs} + \left(e_1^v + \dfrac{e_2^v}{2}\right)\Delta t + \left(\xi_1^v(RS_i) + \dfrac{\xi_2^v(RS_i)}{2}\right)\Delta t \end{cases} \quad (13)$$

$$\begin{cases} \cdots \\ \cdots \end{cases}$$

$$\begin{cases} v_k^{sim}(RS_i) = v_k^{obs} + e_k^v + \xi_k^v(RS_i) \\ x_k^{sim}(RS_i) = x_k^{obs} + \left(\displaystyle\sum_{j=1}^{k-1} e_j^v + \dfrac{e_k^v}{2}\right)\Delta t + \left(\displaystyle\sum_{j=1}^{k-1} \xi_j^v(RS_i) + \dfrac{\xi_k^v(RS_i)}{2}\right)\Delta t \end{cases}$$

Note that $\xi_k$ is a function of $\left(\xi_1(RS_i), \xi_2(RS_i), \cdots, \xi_{k-1}(RS_i)\right)$ so that $\xi_k$ is not an independent and identically distributed variable. The variance of $\xi_k$ is different among different time step, which can be described as $\sigma_k^2$. Given that $\xi_k^v(RS_i)$ is just one sample from the distribution of $\xi_k$, the mean squared error on velocity is:

$$MSE_K^v = \sum_{k=1}^{K}(v_k^{sim} - v_k^{obs})^2 = \sum_{k=1}^{K}\left(e_k^v + \xi_k^v\right)^2 = \sum_{k=1}^{K}\left(\left(e_k^v\right)^2 + \left(\xi_k^v(RS_i)\right)^2 + 2e_k^v\xi_k^v(RS_i)\right) \quad (14)$$

### *4.2 The mechanism of MRMean on errors*

If we use the *RMSE* as GoF and the velocity as MoP, the objective function of the *MRMean* can be written as:

$$\begin{aligned} z &= \operatorname*{mean}_{i\in[1,\ldots,N]}\{MSE_K^V\} = \operatorname*{mean}_{i\in[1,\ldots,N]}\left\{\sum_{k=1}^{K}\left(\left(e_k^v\right)^2 + \left(\xi_k^v(RS_i)\right)^2 + 2e_k^v\xi_k^v(RS_i)\right)\right\} \\ &= \operatorname*{mean}_{i\in[1,\ldots,N]}\left\{e^T e + \xi^T(RS_i)\xi(RS_i) + 2e^T\xi(RS_i)\right\} \end{aligned} \quad (15)$$

here $e = (e_1^v, e_2^v, \cdots, e_K^v)^T$, $\xi(RS_i) = (\xi_1^v(RS_i), \xi_2^v(RS_i), \cdots, \xi_K^v(RS_i))^T$ and $\sigma = (\sigma_1, \sigma_2, \cdots, \sigma_K)^T$. Given that the expectation of $\xi_k^v(RS_{i_k})$ is zero and $e_k^v$ is a fixed value under one set of parameters, one has

$$\operatorname*{mean}_{i \in [1,\ldots,N]} \left\{ 2e^T \xi(RS_i) \right\} = 0 \tag{16}$$

For simplicity, we assume that $\xi_k^v(RS_i)$ follows the normal distribution $N(0, \sigma_k^2)$ so that the second item

$$\operatorname*{mean}_{i \in [1,\ldots,N]} \left\{ \xi^T(RS_i)\xi(RS_i) \right\} = \sum_{k=1}^{K} E_i\left((\xi_k^v(RS_i))^2\right) = \sum_{k=1}^{K} \left( E_i^2\left(\xi_k^v(RS_i)\right) + \sigma_k^2 \right) = \sigma^T \sigma \tag{17}$$

Therefore

$$z = \operatorname*{mean}_{i \in [1,\ldots,N]} \{MSE_K^V\} = e^T e + \sigma^T \sigma \tag{18}$$

In fact, the objective function of *MRMean* is just a combination of the deterministic errors and the variance of stochastic errors. When minimizing $z$ to find the optimal solution, it will obviously minimize the deterministic error $e^T e$, and the variance $\sigma^T \sigma$ tends to zero at the same time. However, when the value of $\sigma_k^2$ tends to zero, the stochastic CF models will degrade into a deterministic one which explains the results of *MRMean* in sections 3.2 and 3.2. The mechanism still holds when adopting spacing as MoP.

### *4.3 The mechanism of MRMin on errors*

As for the *MRMin*, the objective function can be written as

$$z = \min_{i \in [1,\ldots,N]} \{MSE_K^V\} = \min_{i \in [1,\ldots,N]} \left\{ e^T e + \xi^T(RS_i)\xi(RS_i) + 2e^T \xi(RS_i) \right\} \tag{19}$$

Hence, minimizing the objective function can be reformulated as

$$\min z = \min \left\{ e^T e + \min_{i \in [1,\ldots,N]} \left\{ \xi(RS_i)^T \xi(RS_i) + 2e^T \xi(RS_i) \right\} \right\} \tag{20}$$

When calibrating the stochastic CF models, there are two goals. (a) Minimize the parameter epistemic error which has been included as $e^T e$ in the objective function; (b) Keep the aleatoric uncertainty, i.e., let the model keep stochastic rather than deterministic. $RS_i$ can be regarded as another exogenous parameter determined by simulation. Hence, if the objective function of calibrating deterministic CF models is adopted, the effects of random seeds will be ignored and the two error components will be mixed. This means that the calibration of stochastic CF models should be an optimization model with two steps to dispose of the two error components separately. The *MRMin* method converts the calibration into a nested optimization model. The first step of the optimization model is as follows:

$$\begin{cases} \min_{i \in [1,...,N]} f(e(\beta), \xi(\beta, RS_i)) \\ RS^* = \arg\min_{RS_i} f(e(\beta), \xi(\beta, RS_i)) = \arg\min_{RS_i} \left\{ \xi(\beta, RS)^T \xi(\beta, RS) + 2e^T \xi(\beta, RS) \right\} \end{cases} \quad (21)$$

The logic of the first (inner) level is easy to understand. It finds the optimal RS under which the realization is most consistent with reality. From the perspective of driving behavior, it aims to find the trajectory that is most similar to the observed trajectory. This is achieved under the fixed parameters, which optimize the stochasticity-related errors $\xi^T \xi + 2e^T \xi$. Meanwhile, the solution of the first step of calibration does not directly involve the property of stochastic error itself. In other words, the optimization objective does not directly restrict the size of stochasticity. Thus, it does not affect the calibration of the stochastic parameters as the *MRMean* does.

In the second (outer) level of the optimization model, the aim is to minimize the error of parameter epistemic uncertainty.

$$\begin{cases} \min_{\beta} F(e(\beta), \xi(\beta, RS^*)) \\ \beta^* = \arg\min_{\beta} F(e(\beta), \xi(\beta, RS^*)) = \arg\min_{e} \left\{ e^T e + \xi^T(\beta, RS^*)\xi(\beta, RS^*) + 2e^T \xi(\beta, RS^*) \right\} \end{cases} \quad (22)$$

The calibration for deterministic CF models can be treated as an optimization model only with the second step. It only aims to minimize the parameter epistemic error without dealing with the stochastic error. Therefore, it can be directly expressed as solving a single one-step objective function. On the other hand, the first step of the nested optimization model is equivalent to another special calibration scene. For instance, the parameters are iterated to approach the preset values in the solving process of calibrating one synthetic trajectory. In this case, the epistemic error is zero. Thus, the criterion to end the iteration is that the same simulated trajectory can be found in multiple simulations. As long as the number of simulations is large enough, it is capable of finding such a trajectory that has zero error with the synthetic trajectory. The summary of different calibration frameworks is shown in Table. 9.

Table. 9. Optimization framework and solution properties for different calibration scenarios

| Objective Function | Deterministic CF models | Stochastic CF models |
|---|---|---|
| Experimental data | $F(e(\beta^*)) > 0$ | $\begin{cases} F(e(\beta^*), \xi(\beta^*, RS^*)) > 0 \\ f(e(\beta), \xi(\beta, RS^*)) = 0 \end{cases}$ |
| Synthetic data | $F(e(\beta^*)) = 0$ | $\begin{cases} F(e(\beta^*), \xi(\beta^*, RS^*)) = 0 \\ \lim_{N \to \infty} f(e(\beta), \xi(\beta, RS^*)) = 0 \end{cases}$ |
| Decision Variables | $\beta$ | $\beta$ and $RS$ |
| Error Components | $e$ | $e$ and $\xi$ |

## 5 Selection of MoP

The commonly used MoP includes velocity and spacing. By conducting the error propagation analysis, Punzo and Montanino (2016) confirmed that the errors in the spacing are cumulative on velocity so that it contains more degrees of freedom. Hence, when the spacing is selected as MoP, the objective function can keep the memory of model residuals occurrence times. This issue is of great importance (Punzo and Montanino, 2016) because there is only one global solution for the calibration of deterministic CF models. However, when it comes to the adoption of MoP for stochastic CF models, the errors in spacing may still include many degrees of freedom as the errors in velocity in the form of *MSE*. Actually, the errors on spacing also contain the stochastic error items of the aleatoric uncertainty and the calibration process has been converted from a one-level optimization model into a nested optimization model. Thus, the property of objective function has greatly changed and the role of the MoP needs to be revisited.

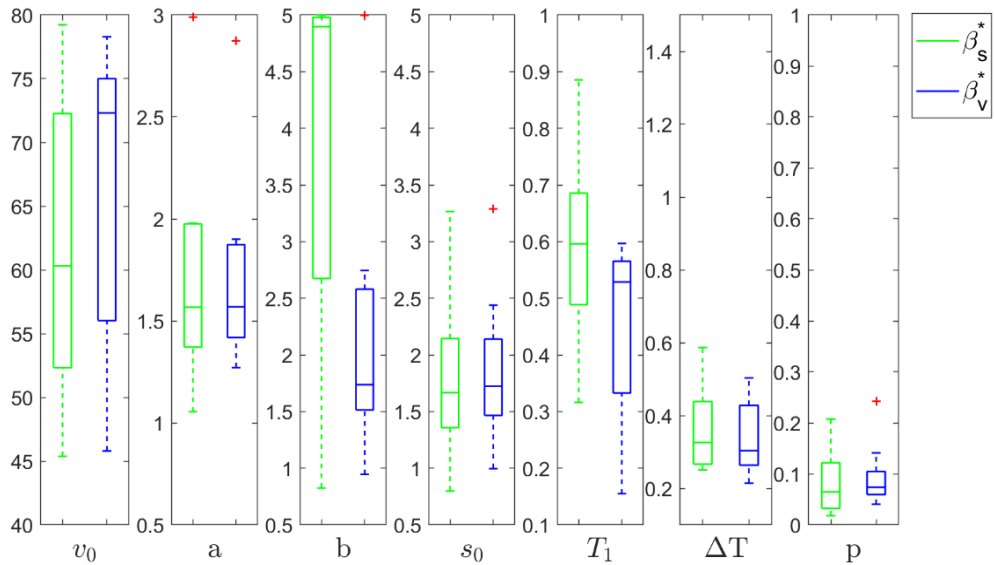

Fig. 10. Boxplot of calibration results for experimental trajectories of 2D-IDM by *MRMin* on spacing (in green) and velocity (in blue), respectively. The calibration boundaries are presented in Table. 4.

Fig. 10 shows the distribution of calibrated parameters sets on spacing and velocity. The values of most of these parameters are similar for the two MoPs. Fig. 11 presents the errors in spacing and velocity for two calibrated parameter sets with different MoPs. With the same 500 random seeds, the parameter set calibrated on spacing may not always outperform the parameter set calibrated on velocity under the stochastic calibration framework. As can be seen in Fig. 11, there are three kinds of relations of calibration results on the two MoPs. Sometimes the parameter set calibrated on spacing performs better, see Fig. 11(a). The parameter set calibrated on velocity could also perform better, see Fig. 11(b). In addition, they can also show similar performance, see Fig. 11(c). Hence, from the perspective of calibration on stochastic CF models, there seems to be no obvious priority between them.

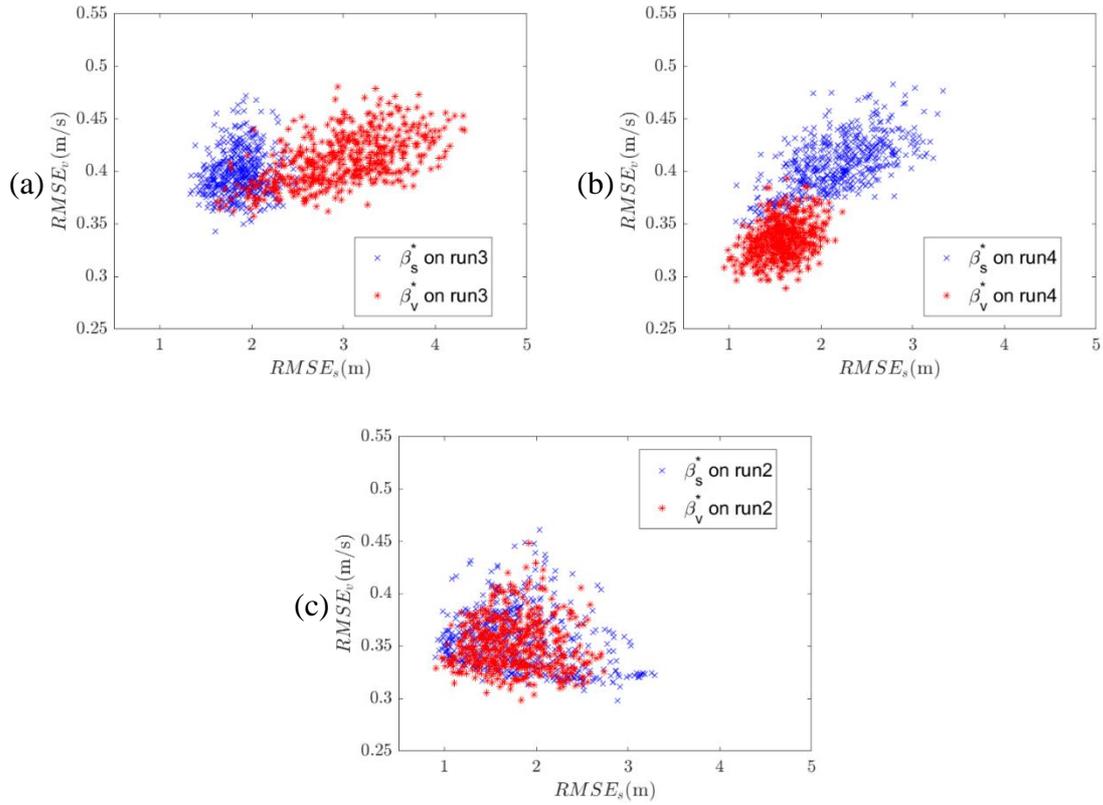

Fig. 11. The errors calculated with RMSE under the coordinate axis on spacing and velocity. The 500 blue points are simulated by using the parameter set calibrated by MRMin on spacing ($\beta_s^*$) with the experimental trajectory of different runs (a)Run 3 (b)Run 4 (c)Run 2. The 500 red points are simulated by using the parameter set calibrated by MRMin on velocity ($\beta_v^*$) with the same experimental trajectory. The blue and red points are using the same total of 500 random seeds. Hence, each blue or red point corresponds to one simulation result under one unique random seed.

To further verify this finding, we conduct the error analysis. According to equation (12), the MSE on spacing is the accumulation of the MSE on velocity:

$$MSE_K^x = \sum_{k=1}^{K}(x_k^{sim} - x_k^{obs})^2 = \sum_{k=1}^{K}\left(\left(\sum_{j=1}^{k-1}e_j^v + \frac{e_k^v}{2}\right)\Delta t + \left(\sum_{j=1}^{k-1}\xi_j^v(RS_i) + \frac{\xi_k^v(RS_i)}{2}\right)\Delta t\right)^2$$

$$= \Delta t^2 \sum_{k=1}^{K}\left(\left(\sum_{j=1}^{k-1}e_j^v + \frac{e_k^v}{2}\right) + \left(\sum_{j=1}^{k-1}\xi_j^v(RS_i) + \frac{\xi_k^v(RS_i)}{2}\right)\right)^2$$

$$= \Delta t^2 \sum_{k=1}^{K}\begin{pmatrix}\left(\sum_{j=1}^{k-1}\left(e_j^v + \xi_j^v(RS_i)\right)^2\right) + \frac{\left(e_k^v + \xi_k^v(RS_i)\right)^2}{2} + \\ 2\sum_{m,n\in[1,\ldots,k-1]}\left(e_m^v + \xi_m^v(RS_i)\right)\left(e_n^v + \xi_n^v(RS_i)\right) + \\ \sum_{n\in[1,\ldots,k-1]}\left(e_k^v + \xi_k^v(RS_i)\right)\left(e_n^v + \xi_n^v(RS_i)\right)\end{pmatrix} \quad (23)$$

$$= \Delta t^2 \begin{pmatrix}\frac{1}{4}MSE^v + \sum_{k=1}^{K-1}(K-k)\left(e_k^v + \xi_k^v(RS_i)\right)^2 + \\ 2\sum_{k=1}^{K-1}\left(e_k^v + \xi_k^v(RS_i)\right)\sum_{i=k+1}^{K}(K-i+\frac{1}{2})\left(e_k^v + \xi_k^v(RS_i)\right)\end{pmatrix}$$

Hence, the objective function of the *MRMin* on spacing is as follows:

$$z = \min_{i\in[1,\ldots,N]}\{MSE_K^x\}$$

$$= \Delta t^2 \cdot \min_{i\in[1,\ldots,N]}\begin{pmatrix}\frac{1}{4}MSE^v + \sum_{k=1}^{K-1}(K-k)\left(e_k^v + \xi_k^v(RS_i)\right)^2 + \\ 2\sum_{k=1}^{K-1}\left(e_k^v + \xi_k^v(RS_i)\right)\sum_{i=k+1}^{K}(K-i+\frac{1}{2})\left(e_k^v + \xi_k^v(RS_i)\right)\end{pmatrix}$$

$$= \Delta t^2 \begin{pmatrix}(K+\frac{1}{4})\min_{i\in[1,\ldots,N]}MSE^v + \\ \min_{i\in[1,\ldots,N]}\left\{\begin{array}{l}\sum_{k=1}^{K-1}(K-k)\left(e_k^v + \xi_k^v(RS_i)\right)^2 + \\ 2\sum_{k=1}^{K-1}\left(e_k^v + \xi_k^v(RS_i)\right)\sum_{i=k+1}^{K}(K-i+\frac{1}{2})\left(e_k^v + \xi_k^v(RS_i)\right)\end{array}\right\}\end{pmatrix} \quad (24)$$

$$= \Delta t^2 \begin{pmatrix}(K+\frac{1}{4})\left\{\sum_{k=1}^{K}(e_k^v)^2 + \underline{\min_{i\in[1,\ldots,N]}\left\{\sum_{k=1}^{K}\left(\left(\xi_k^v(RS_i)\right)^2 + 2e_k^v\xi_k^v(RS_i)\right)\right\}}\right\} + \\ \sum_{k=1}^{K-1}(K-k)\left(\left(e_k^v\right)^2 + \underline{\min_{i\in[1,\ldots,N]}\left\{\left(\xi_k^v(RS_i)\right)^2 + 2e_k^v\xi_k^v(RS_i)\right\}}\right) + \\ \underline{\min_{i\in[1,\ldots,N]}\left\{2\sum_{k=1}^{K-1}\left(e_k^v + \xi_k^v(RS_i)\right)\sum_{i=k+1}^{K}(K-i+\frac{1}{2})\left(e_k^v + \xi_k^v(RS_i)\right)\right\}}\end{pmatrix}$$

From the formula, the calibration on spacing is one exactly nested optimization model whose outer level and inner level are different from the calibration on velocity (The underline in Equation (24) represents the inner level for *MRMin-MSE$_s$*). Hence, they are in a more complex relationship than the cumulative one. Therefore, when we adopt the spacing as MoP with *MRMin*, it is not equivalent to minimizing the accumulation of errors on the velocity with *MRMin*. Their calibration results correspond to two different random seeds. Hence, they do not have a one-to-one relationship.

We also validate the calibrated parameters on spacing and velocity by calculating the validation error *MRMin* on spacing (or velocity). We compare the experimental trajectories and the simulated ones, the parameters of which are calibrated by *MRMin* on velocity (or spacing),

$$ValiError_s^v = \min_{i \in [1,...,N]} RMSE(v^{obs}, v_i^{sim}(\beta^*(\min RMSE_s))) \tag{25}$$

where $\beta^*(\min RMSE(s))$ are the optimal parameters calibrated by *MRMin-RMSE* on spacing. Hence, four 9*9 matrices of validation errors can be computed, i.e., $ValiError_v^s$, $ValiError_s^s$, $ValiError_s^v$ and $ValiError_v^v$. For instance, Table. 10 shows the matrix of $ValiError_s^s$. The details of other validation errors can be found in Appendix B.

Table. 10. The cross errors $ValiError_s^v$ of 2D-IDM: validation errors calculated by $MRMin_v$ with parameter sets of different trajectories calibrated by $MRMin_s$.

| $ValiError_s^v$ | | Cross errors calculated on trajectory No.* (Unit: m/s) | | | | | | | | | | |
|---|---|---|---|---|---|---|---|---|---|---|---|---|
| | | 1 | 2 | 3 | 4 | 5 | 6 | 7 | 8 | 9 | Mean | STD |
| Calibrated parameter set of trajectory No.* | 1 | 0.294 | 0.312 | 0.346 | 0.273 | 0.327 | 0.309 | 0.288 | 0.289 | 0.353 | 0.31 | 0.028 |
| | 2 | 0.306 | 0.301 | 0.339 | 0.275 | 0.293 | 0.311 | 0.258 | 0.311 | 0.309 | 0.3 | 0.023 |
| | 3 | 0.336 | 0.288 | 0.262 | 0.278 | 0.284 | 0.286 | 0.253 | 0.306 | 0.286 | 0.287 | 0.024 |
| | 4 | 0.311 | 0.286 | 0.324 | 0.223 | 0.288 | 0.313 | 0.295 | 0.285 | 0.291 | 0.291 | 0.029 |
| | 5 | 0.316 | 0.28 | 0.316 | 0.256 | 0.258 | 0.274 | 0.246 | 0.285 | 0.285 | 0.28 | 0.025 |
| | 6 | 0.331 | 0.288 | 0.28 | 0.282 | 0.259 | 0.231 | 0.244 | 0.302 | 0.262 | 0.275 | 0.03 |
| | 7 | 0.322 | 0.273 | 0.278 | 0.267 | 0.241 | 0.269 | 0.231 | 0.27 | 0.299 | 0.272 | 0.027 |
| | 8 | 0.332 | 0.3 | 0.377 | 0.283 | 0.303 | 0.345 | 0.313 | 0.302 | 0.319 | 0.319 | 0.028 |
| | 9 | 0.324 | 0.31 | 0.321 | 0.294 | 0.306 | 0.293 | 0.284 | 0.307 | 0.238 | 0.297 | 0.026 |
| | | Mean value of All | | | | | | | | | **0.292** | **0.027** |

Table. 11. The cross-validation errors comparison of 2D-IDM calculated with parameter sets of different trajectories calibrated by different MoPs.

| | Validation errors calculated on trajectory No.* | | | | | | | | | Mean of all | STD of all |
|---|---|---|---|---|---|---|---|---|---|---|---|
| | 1 | 2 | 3 | 4 | 5 | 6 | 7 | 8 | 9 | | |
| $ValiError_s^v$ | 0.310 | 0.300 | 0.287 | 0.291 | 0.280 | 0.275 | 0.272 | 0.319 | 0.297 | **0.292** | **0.027** |
| $ValiError_v^v$ | 0.293 | 0.266 | 0.28 | 0.267 | 0.276 | 0.287 | 0.27 | 0.261 | 0.334 | **0.282** | **0.034** |
| $ValiError_s^v$ | 1.474 | 1.093 | 1.349 | 1.351 | 2.565 | 1.974 | 2.339 | 1.049 | 1.553 | **1.639** | **0.552** |
| $ValiError_s^s$ | 1.388 | 1.05 | 1.148 | 1.483 | 1.276 | 1.25 | 1.921 | 1.204 | 1.224 | **1.327** | **0.475** |

The mean values reflect the validation performance of one calibrated set of parameters, see Table 11. The significance test is conducted to explore the difference in the validation performance of different MoPs. We carried out the KS-test and t-test, and the results demonstrate whether we use spacing or velocity as the MoP, the validation performance of the calibrated parameters has no significant difference, see Table. 12. In fact, the difference between 1.327 and 1.639 is only 0.3 m in spacing, which is very small in reality. To summarize, there is no difference in using these two types of MoPs when the stochastic CF models are applied to reality. This is reasonable given that the relation of spacing and velocity is not one-to-one. Together with the results in Fig. 11, it is recommended that when calibrating stochastic CF models, both spacing and velocity are adopted. Then the performance of calibration results of the two MoPs should be evaluated. The parameter set with better performance should be selected as the calibration results.

Table. 12. The significant test of the validation errors of 2D-IDM on velocity and spacing

| Validation Errors | Averaged validation *MRMin* on spacing | Averaged validation *MRMin* on velocity |
|---|---|---|
| Parameters calibrated by *MRMin* on spacing | 1.327 | 0.292 |
| Parameters calibrated by *MRMin* on velocity | 1.639 | 0.282 |
| p-value in t-test | 0.13554 | 0.25542 |
| p-value in KS-test | 0.2500 | 0.2500 |

## 6 Conclusions and future directions

Although remarkable progress has been made in the field of traffic flow model calibration, all these were made under the deterministic framework. The calibration study of stochastic CF models is lacking. The stochasticity adds noises to the calibration process and is not straightforward to deal with. This paper conducts a detailed investigation of the calibration methodology of stochastic CF models by using a new dataset of the CF experiment. Our analysis reveals that (i) the stochasticity in driving behavior can be captured by one set of parameters with stochastic CF models. (ii) the *MRMin-RMSE* should be adopted as the objective function because the formulation of *MRMin-RMSE* effectively separates the aleatoric uncertainty and epistemic uncertainty and turns the calibration process into a nested optimization problem. (iii) there is no priority between spacing and velocity to be adopted as MoP under the stochastic framework of calibration.

These findings are different and novel from the previous studies on the calibration of deterministic CF models. Hence, we establish the new calibration framework for the stochastic CF models. The effective calibration methodology of stochastic CF models can also promote its application because the individual driving behavior stochasticity under the same driving environment can be captured by one set of parameters.

To summarize, the main target when calibrating stochastic CF models is to minimize the epistemic uncertainty by considering the impact of aleatoric uncertainty rising from the stochastic items of models. In the future, more objective function forms and methods under the calibration framework of stochastic CF models can be developed. Meanwhile, another goal to pursue is to decrease the computational complexity and optimization solution time, which requires a more high-efficiency global optimization algorithm.

## Acknowledgment

This work is supported by the National Natural Science Foundation of China (Grant No. 72222021, 71931002, 72010107004, 72288101).

## Appendix A

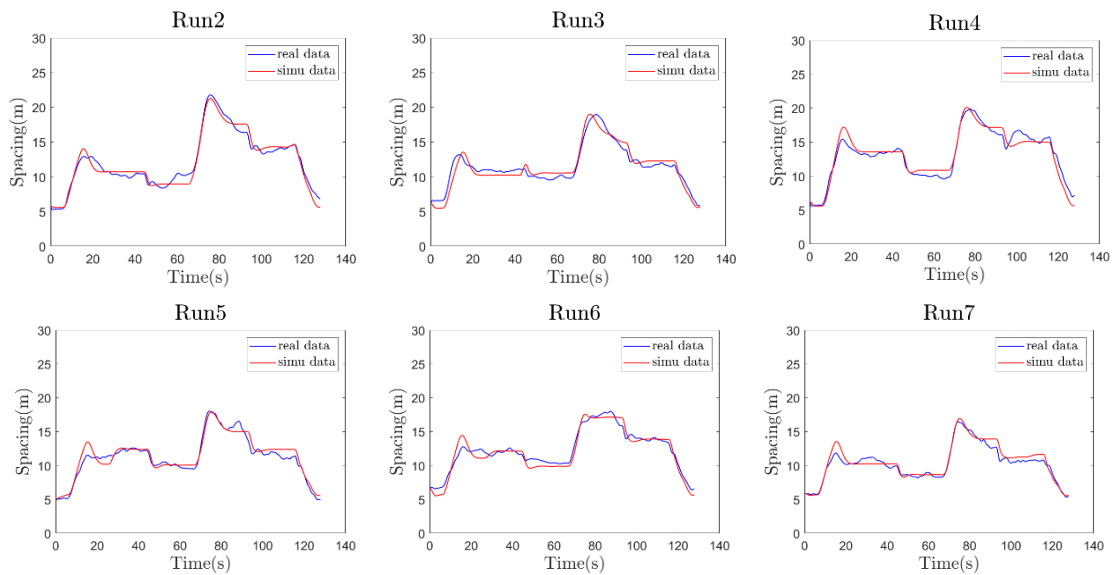

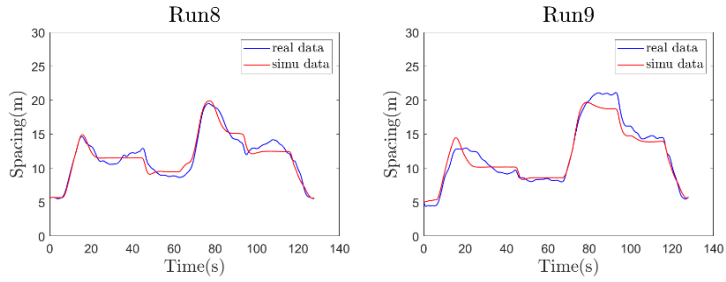

Fig. 12. The spacing profile of the best realization of simulated trajectories with one set of parameters calibrated by *MRMin* of 2D-IDM.

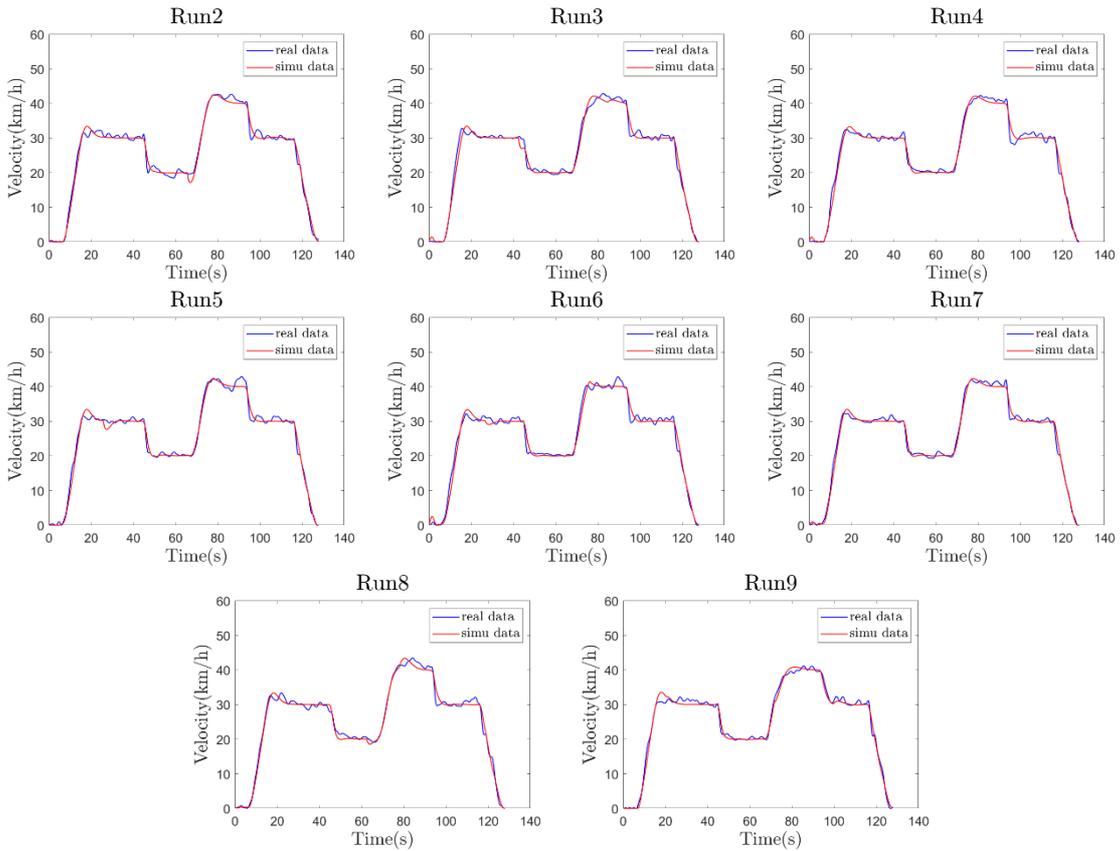

Fig. 13. The velocity profile of the simulated trajectories with one set of parameters calibrated by *MRMin* of 2D-IDM.

Table. 13. The validation errors of 2D-IDM calculated by *MRMin* with parameter sets of different trajectories calibrated by *MRMean*. The red value represents the averaged validation performance of the parameter set calibrated with one trajectory on the other 8 trajectories.

| Validation Error | | Validation calculated on trajectory No.* (Units: m) | | | | | | | | | | |
|---|---|---|---|---|---|---|---|---|---|---|---|---|
| | | 1 | 2 | 3 | 4 | 5 | 6 | 7 | 8 | 9 | Mean | STD |
| Calibrated parameter set of trajectory No.* | 1 | 0.87 | 1.968 | 1.759 | 1.346 | 1.852 | 1.223 | 2.602 | 1.439 | 2.538 | 1.733 | 0.582 |
| | 2 | 1.201 | 1.539 | 1.389 | 1.583 | 1.336 | 1.094 | 2.221 | 1.18 | 2.024 | 1.507 | 0.387 |
| | 3 | 1.69 | 1.851 | 0.763 | 2.249 | 0.995 | 1.272 | 1.285 | 1.193 | 2.536 | 1.537 | 0.588 |
| | 4 | 1.103 | 2.015 | 2.15 | 1.081 | 2.061 | 1.307 | 3.071 | 1.786 | 2.563 | 1.904 | 0.669 |
| | 5 | 1.735 | 1.928 | 1.003 | 2.232 | 0.716 | 1.206 | 1.211 | 1.184 | 2.482 | 1.522 | 0.599 |
| | 6 | 1.309 | 1.766 | 1.38 | 1.601 | 1.278 | 0.639 | 2.144 | 1.399 | 2.193 | 1.523 | 0.477 |
| | 7 | 2.645 | 2.647 | 1.455 | 3.223 | 1.327 | 2.167 | 0.548 | 1.888 | 3.251 | 2.128 | 0.911 |
| | 8 | 1.287 | 1.707 | 1.128 | 1.828 | 1.027 | 1.222 | 1.714 | 0.924 | 2.471 | 1.479 | 0.494 |
| | 9 | 2.336 | 2.429 | 2.677 | 2.499 | 2.585 | 1.973 | 3.221 | 2.527 | 1.062 | 2.368 | 0.589 |
| | | Mean value of All | | | | | | | | | **1.745** | **0.588** |

Table. 14. The validation errors of 2D-IDM calculated by *MRMin* with parameter sets of different trajectories calibrated by *MLE*. The red value represents the averaged validation performance of the parameter set calibrated with one trajectory on all the trajectories.

| Validation Error | | Validation calculated on trajectory No.* (Unit:m) | | | | | | | | | | |
|---|---|---|---|---|---|---|---|---|---|---|---|---|
| | | 1 | 2 | 3 | 4 | 5 | 6 | 7 | 8 | 9 | Mean | STD |
| Calibrated parameter set of trajectory No.* | 1 | 0.757 | 1.707 | 1.467 | 1.136 | 1.605 | 1.086 | 2.208 | 1.263 | 2.366 | 1.511 | 0.527 |
| | 2 | 1.032 | 1.254 | 0.766 | 1.242 | 0.86 | 0.773 | 1.319 | 0.942 | 1.64 | 1.092 | 0.294 |
| | 3 | 1.529 | 1.733 | 0.682 | 1.95 | 1.05 | 1.148 | 1.245 | 1.216 | 2.441 | 1.444 | 0.531 |
| | 4 | 0.997 | 1.623 | 1.61 | 1.002 | 1.485 | 0.875 | 2.378 | 1.406 | 2.164 | 1.505 | 0.517 |
| | 5 | 1.673 | 1.884 | 1.058 | 2.155 | 0.676 | 1.159 | 1.207 | 1.11 | 2.464 | 1.487 | 0.588 |
| | 6 | 1.13 | 1.56 | 1.268 | 1.342 | 1.111 | 0.624 | 1.912 | 1.215 | 1.92 | 1.343 | 0.409 |
| | 7 | 2.564 | 2.55 | 1.365 | 3.132 | 1.241 | 2.099 | 0.529 | 1.817 | 3.187 | 2.054 | 0.899 |
| | 8 | 1.165 | 1.545 | 1.08 | 1.645 | 0.978 | 1.084 | 1.51 | 0.85 | 2.264 | 1.347 | 0.441 |
| | 9 | 1.665 | 1.932 | 1.882 | 1.751 | 1.67 | 1.488 | 2.461 | 1.949 | 0.894 | 1.744 | 0.42 |
| | | Mean value of All | | | | | | | | | **1.503** | **0.514** |

**Appendix B**

Table. 15. The cross errors $ValiError_v^v$ of 2D-IDM: validation errors calculated by $MRMin_v$ with parameter sets of different trajectories calibrated by $MRMin_v$.

| $ValiError_v^v$ | | Cross errors calculated on trajectory No.* (Unit:m/s) | | | | | | | | | | |
|---|---|---|---|---|---|---|---|---|---|---|---|---|
| | | 1 | 2 | 3 | 4 | 5 | 6 | 7 | 8 | 9 | Mean | STD |
| Calibrated parameter set of trajectory No.* | 1 | 0.289 | 0.298 | 0.336 | 0.252 | 0.302 | 0.31 | 0.28 | 0.271 | 0.303 | 0.293 | 0.024 |
| | 2 | 0.315 | 0.282 | 0.27 | 0.267 | 0.244 | 0.253 | 0.232 | 0.265 | 0.263 | 0.266 | 0.024 |
| | 3 | 0.344 | 0.313 | 0.214 | 0.289 | 0.278 | 0.229 | 0.241 | 0.307 | 0.301 | 0.28 | 0.043 |
| | 4 | 0.299 | 0.278 | 0.282 | 0.252 | 0.253 | 0.233 | 0.239 | 0.275 | 0.297 | 0.267 | 0.024 |
| | 5 | 0.326 | 0.286 | 0.276 | 0.281 | 0.23 | 0.246 | 0.219 | 0.309 | 0.313 | 0.276 | 0.038 |

|   | 6 | 0.346 | 0.313 | 0.276 | 0.304 | 0.252 | 0.217 | 0.236 | 0.322 | 0.319 | 0.287 | 0.044 |
|   | 7 | 0.329 | 0.28 | 0.241 | 0.276 | 0.237 | 0.233 | 0.203 | 0.308 | 0.327 | 0.27 | 0.045 |
|   | 8 | 0.298 | 0.27 | 0.248 | 0.263 | 0.256 | 0.253 | 0.221 | 0.256 | 0.284 | 0.261 | 0.022 |
|   | 9 | 0.374 | 0.334 | 0.363 | 0.349 | 0.328 | 0.342 | 0.352 | 0.347 | 0.217 | 0.334 | 0.046 |
|   |   |   |   |   |   | Mean value of All |   |   |   |   | 0.282 | 0.034 |

Table. 16. The cross errors $ValiError_v^s$ of 2D-IDM: validation errors calculated by $MRMin_s$ with parameter sets of different trajectories calibrated by $MRMin_v$.

| $ValiError_v^s$ |   | Cross errors calculated on trajectory No.* (Unit:m) |   |   |   |   |   |   |   |   |   |   |
|---|---|---|---|---|---|---|---|---|---|---|---|---|
|   |   | 1 | 2 | 3 | 4 | 5 | 6 | 7 | 8 | 9 | Mean | STD |
| Calibrated parameter set of trajectory No.* | 1 | 1.218 | 1.921 | 1.085 | 1.775 | 1.019 | 1.189 | 1.464 | 1.066 | 2.53 | 1.474 | 0.51 |
|   | 2 | 1.047 | 1.184 | 0.873 | 1.458 | 0.862 | 0.752 | 0.781 | 1.069 | 1.815 | 1.093 | 0.35 |
|   | 3 | 1.261 | 1.576 | 1.135 | 1.352 | 1.242 | 0.763 | 1.102 | 1.417 | 2.295 | 1.349 | 0.421 |
|   | 4 | 1.145 | 1.677 | 1.067 | 1.54 | 0.964 | 0.711 | 1.527 | 1.155 | 2.376 | 1.351 | 0.493 |
|   | 5 | 3.143 | 3.098 | 1.972 | 3.772 | 1.659 | 2.75 | 0.832 | 2.408 | 3.453 | 2.565 | 0.942 |
|   | 6 | 2.329 | 2.433 | 1.468 | 2.641 | 1.222 | 1.813 | 0.807 | 1.878 | 3.171 | 1.974 | 0.744 |
|   | 7 | 2.92 | 3.001 | 1.775 | 3.199 | 1.419 | 2.406 | 0.734 | 2.132 | 3.461 | 2.339 | 0.906 |
|   | 8 | 1.121 | 0.992 | 1.033 | 1.15 | 0.843 | 0.761 | 0.736 | 1.083 | 1.723 | 1.049 | 0.295 |
|   | 9 | 1.405 | 1.618 | 1.631 | 1.769 | 1.532 | 1.162 | 2.14 | 1.576 | 1.149 | 1.553 | 0.304 |
|   |   |   |   |   |   | Mean value of All |   |   |   |   | **1.639** | **0.552** |

Table. 17. The cross errors $ValiError_s^s$ of 2D-IDM: validation errors calculated by $MRMin_s$ with parameter sets of different trajectories calibrated by $MRMin_s$.

| $ValiError_s^s$ |   | Cross errors calculated on trajectory No.* (Unit:m) |   |   |   |   |   |   |   |   |   |   |
|---|---|---|---|---|---|---|---|---|---|---|---|---|
|   |   | 1 | 2 | 3 | 4 | 5 | 6 | 7 | 8 | 9 | Mean | STD |
| Calibrated parameter set of trajectory No.* | 1 | 0.681 | 1.719 | 1.444 | 1.112 | 1.469 | 1.05 | 1.629 | 1.151 | 2.237 | 1.388 | 0.453 |
|   | 2 | 1.022 | 0.883 | 0.944 | 1.03 | 1.039 | 0.945 | 0.841 | 0.884 | 1.859 | 1.05 | 0.312 |
|   | 3 | 1.209 | 1.323 | 0.55 | 1.499 | 0.929 | 0.744 | 0.972 | 1.057 | 2.05 | 1.148 | 0.444 |
|   | 4 | 0.956 | 1.529 | 1.53 | 1.017 | 1.666 | 0.95 | 2.29 | 1.224 | 2.185 | 1.483 | 0.503 |
|   | 5 | 1.424 | 1.548 | 1.014 | 1.855 | 0.584 | 0.89 | 0.775 | 1.128 | 2.268 | 1.276 | 0.546 |
|   | 6 | 1.147 | 1.699 | 1.18 | 1.287 | 0.908 | 0.441 | 1.286 | 1.221 | 2.079 | 1.25 | 0.458 |
|   | 7 | 2.441 | 2.455 | 1.151 | 2.922 | 1.076 | 1.944 | 0.513 | 1.712 | 3.074 | 1.921 | 0.882 |
|   | 8 | 0.958 | 1.396 | 1.109 | 1.383 | 0.926 | 1 | 1.107 | 0.712 | 2.242 | 1.204 | 0.445 |
|   | 9 | 1.242 | 1.486 | 1.421 | 1.545 | 0.992 | 1.093 | 1.05 | 1.3 | 0.891 | 1.224 | 0.232 |
|   |   |   |   |   |   | Mean value of All |   |   |   |   | **1.327** | **0.475** |

**References**

Bando, M., Hasebe, K., Nakayama, A., Shibata, A., Sugiyama, Y., 1995. Dynamical model of traffic congestion and numerical simulation. Phys. Rev. E 51, 1035-1042.

Chen, D., Laval, J.A., Ahn, S., Zheng, Z., 2012. Microscopic traffic hysteresis in traffic oscillations: A behavioral perspective. Transp. Res. Part B Methodol. 46, 1440-1453.

Ciuffo, B., Punzo, V., 2014. "No Free Lunch" Theorems Applied to the Calibration of Traffic Simulation Models. IEEE Trans. Intell. Transp. Syst. 15, 553-562.

Hoogendoorn, S., Hoogendoorn, R., 2010. Calibration of microscopic traffic-flow models using multiple data sources. Philos. Trans. R. Soc. Math. Phys. Eng. Sci. 368, 4497-4517.

Jiang, R., Hu, M.-B., Zhang, H.M., Gao, Z.-Y., Jia, B., Wu, Q.-S., 2015. On some experimental features of car-following behavior and how to model them. Transp. Res. Part B Methodol. 80, 338-354.

Jiang, R., Hu, M.-B., Zhang, H.M., Gao, Z.-Y., Jia, B., Wu, Q.-S., Wang, B., Yang, M., 2014. Traffic Experiment Reveals the Nature of Car-Following. PLoS ONE 9, e94351.

Jiang, R., Wu, Q., Zhu, Z., 2001. Full velocity difference model for a car-following theory. Phys. Rev. E 64, 017101.

Krajewski, R., Bock, J., Kloeker, L., Eckstein, L., 2018. The highD Dataset: A Drone Dataset of Naturalistic Vehicle Trajectories on German Highways for Validation of Highly Automated Driving Systems, in: 2018 21st International Conference on Intelligent Transportation Systems (ITSC). Presented at the 2018 21st International Conference on Intelligent Transportation Systems (ITSC), pp. 2118-2125.

Laval, J.A., Toth, C.S., Zhou, Y., 2014. A parsimonious model for the formation of oscillations in car-following models. Transp. Res. Part B Methodol. 70, 228-238.

Lee, J.-B., Ozbay, K., 2009. New Calibration Methodology for Microscopic Traffic Simulation Using Enhanced Simultaneous Perturbation Stochastic Approximation Approach. Transp. Res. Rec. J. Transp. Res. Board 2124, 233-240.

Lee, S., Ngoduy, D., Keyvan-Ekbatani, M., 2019. Integrated deep learning and stochastic car-following model for traffic dynamics on multi-lane freeways. Transp. Res. Part C Emerg. Technol. 106, 360-377.

Lee, S., Ryu, I., Ngoduy, D., Hoang, N.H., Choi, K., 2021. A stochastic behaviour model of a personal mobility under heterogeneous low-carbon traffic flow. Transp. Res. Part C Emerg. Technol. 128, 103163.

Li, L., Chen, X. (Micheal), Zhang, L., 2016. A global optimization algorithm for trajectory data based car-following model calibration. Transp. Res. Part C Emerg. Technol. 68, 311-332.

Newell, G.F., 2002. A simplifed car-following theory: a lower order model. Transp. Res. Part B Methodol. 36, 195-205.

Ngoduy, D., 2021. Noise-induced instability of a class of stochastic higher order continuum traffic models. Transp. Res. Part B Methodol. 150, 260-278.

Ngoduy, D., Lee, S., Treiber, M., Keyvan-Ekbatani, M., Vu, H.L., 2019. Langevin method for a continuous stochastic car-following model and its stability conditions. Transp. Res. Part C Emerg. Technol. 105, 599-610.

Punzo, V., Ciuffo, B., Montanino, M., 2012. Can Results of car-following Model Calibration Based on Trajectory Data be Trusted? Transp. Res. Rec. J. Transp. Res. Board 2315, 11-24.

Punzo, V., Montanino, M., 2020. A two-level probabilistic approach for validation of stochastic traffic simulations: impact of drivers' heterogeneity models. Transp. Res. Part C Emerg. Technol. 121, 102843.

Punzo, V., Montanino, M., 2016. Speed or spacing? Cumulative variables, and convolution of model errors and time in traffic flow models validation and calibration. Transp. Res. Part B Methodol. 91, 21-33.

Punzo, V., Montanino, M., Ciuffo, B., 2015. Do We Really Need to Calibrate All the Parameters? Variance-Based Sensitivity Analysis to Simplify Microscopic Traffic Flow Models. IEEE Trans. Intell. Transp. Syst. 16, 184-193.

Punzo, V., Simonelli, F., 2005. Analysis and Comparison of Microscopic Traffic Flow Models with Real Traffic Microscopic Data. Transp. Res. Rec. 1934, 53-63.

Punzo, V., Zheng, Z., Montanino, M., 2021. About calibration of car-following dynamics of automated and human-driven vehicles: Methodology, guidelines and codes. Transp. Res. Part C Emerg. Technol. 128, 103165.

Sharma, A., Zheng, Z., Bhaskar, A., 2019. Is more always better? The impact of vehicular trajectory completeness on car-following model calibration and validation. Transp. Res. Part B Methodol. 120, 49-75.

Tian, J., Jiang, R., Jia, B., Gao, Z., Ma, S., 2016a. Empirical analysis and simulation of the concave growth pattern of traffic oscillations. Transp. Res. Part B Methodol. 93, 338-354.

Tian, J., Li, G., Treiber, M., Jiang, R., Jia, N., Ma, S., 2016b. Cellular automaton model simulating spatiotemporal patterns, phase transitions and concave growth pattern of oscillations in traffic flow. Transp. Res. Part B Methodol. 93, 560-575.

Tian, J., Zhang, H.M., Treiber, M., Jiang, R., Gao, Z.-Y., Jia, B., 2019. On the role of speed adaptation and spacing indifference in traffic instability: Evidence from car-following experiments and its stochastic model. Transp. Res. Part B Methodol. 129, 334-350.

Treiber, M., Hennecke, A., Helbing, D., 2000. Congested Traffic States in Empirical Observations and Microscopic Simulations. Phys. Rev. E 62, 1805-1824.

Treiber, M., Kanagaraj, V., 2015. Comparing numerical integration schemes for time-continuous car-following models. Phys. Stat. Mech. Its Appl. 419, 183-195.

Treiber, M., Kesting, A., 2018. The Intelligent Driver Model with stochasticity – New insights into traffic flow oscillations. Transp. Res. Part B Methodol. 117, 613-623.

Treiber, M., Kesting, A., 2013. Microscopic Calibration and Validation of Car-Following Models – A Systematic Approach. Procedia - Soc. Behav. Sci. 80, 922-939.

Wang, H., Wang, W., Chen, J., Jing, M., 2010. Using Trajectory Data to Analyze Intradriver Heterogeneity in Car-Following. Transp. Res. Rec. J. Transp. Res. Board 2188, 85-95.

Xu, T., Laval, J., 2020. Statistical inference for two-regime stochastic car-following models. Transp. Res. Part B Methodol. 134, 210-228.

Xu, T., Laval, J.A., 2019. Analysis of a Two-Regime Stochastic Car-Following Model: Explaining Capacity Drop and Oscillation Instabilities. Transp. Res. Rec. J. Transp. Res. Board 2673, 610-619.

Zheng, S.-T., Jiang, R., Tian, J., Li, X., Treiber, M., Li, Z.-H., Gao, L.-D., Jia, B., 2022. Empirical and experimental study on the growth pattern of traffic oscillations upstream of fixed bottleneck and model test. Transp. Res. Part C Emerg. Technol. 140, 103729.

Zhong, R.X., Fu, K.Y., Sumalee, A., Ngoduy, D., Lam, W.H.K., 2016. A cross-entropy method and probabilistic sensitivity analysis framework for calibrating microscopic traffic models. Transp. Res. Part C Emerg. Technol. 63, 147-169.

Zhou, M., Qu, X., Li, X., 2017. A recurrent neural network based microscopic car following model to predict traffic oscillation. Transp. Res. Part C Emerg. Technol. 84, 245-264.